\documentclass[a4paper,twoside]{article}
\newcommand{\affil}[1]{$^{\rm #1}$}
\usepackage[authoryear]{natbib}
\bibpunct{(}{)}{;}{a}{}{,}
\usepackage{graphicx}
\date{} 
\title{\large\bf\flushleft KIC 2557430: A Triple System Containing \\ One $\gamma$ Dor and Two Flaring Components?}
\author{\parbox{\textwidth}{\flushleft
\vspace{-0.5cm}
{\it Ceren Kamil\affil{1}, Hasan Ali Dal\affil{1},\affil{2}}\\
\vspace{0.4cm}
{\small \affil{1}\,Department of Astronomy and Space Sciences, University of Ege, Bornova, 35100 ~\.{I}zmir, Turkey}\\
{\small \affil{2}\,Corresponding Author, Email: ali.dal@ege.edu.tr}}}
\begin{document}
\twocolumn[
\begin{changemargin}{.8cm}{.5cm}
\begin{minipage}{.9\textwidth}
\vspace{-1cm}
\maketitle
%
%
\small{\bf Abstract:} The results obtained from the photometrical models are presented for KIC 2557430, which is known as an eclipsing binary system in the literature. Its light curve was analysed for the first time. The secondary component temperature was found as 6271$\pm$1 K. The mass ratio was computed as 0.868$\pm$0.002, while the inclination ($i$) of the system was found as 69$^\circ$.75$\pm$0$^\circ$.01. There is a third light contribution of 0.339$\%$ in the total light. 50 frequencies were found in the period analysis. 48 frequencies of them are caused due to the primary component, a $\gamma$ Doradus star, while two of them are caused by the rotational modulation due to the cool spots. 69 flares were detected in the analyses. Two OPEA models were derived for these flares detected from KIC 2557430. Obtained OPEA models indicate that the flares were come from two different sources. The Plateau value was found to be 1.4336$\pm$0.1104 s for Source 1, which is seen as possible the secondary component and 0.7550$\pm$0.0677 s for Source 2, which is seen as possible third body. The half-life value was computed as 2278.1 s for Group 1 and 1811.2 s for Group 2. The flare frequency $N_{1}$ was found to be 0.02726 $h^{-1}$ and the flare frequency $N_{2}$ was computed as 0.00002 for Group 1, while $N_{1}$ was found to be 0.01977 $h^{-1}$ and $N_{2}$ was computed as 0.00001 for Group 2. As a result of this study, KIC 2557430 is a possible triple system. Considering both components' temperatures and obtained two different OPEA models, the primary star is a $\gamma$ Doradus-type star, and the secondary component is a chromospherically active star, which has both the cool spot and flare activities, and also the third body, whose the membership status is controversial for this system, is a flare star.\\


\medskip
\medskip
\end{minipage}
\end{changemargin}
]
\small

\section{Introduction}
It is well known for about four decades that sixty-five percent of the stars in our Galaxy is consisted from red drawfs, whose seventy-five percent called UV Ceti type stars exhibits strong flare activity \citep{Rod86}. There are lots of studies in the literature that UV Ceti type stars are very young stars and their population increases in the open cluster and especially in the associations \citep{Mir90, Pig90}. The population rate of the flaring red dwarfs is decreasing while the age of the cluster is increasing, as it is expected according to Skumanich's law \citep{Sku72, Pet91, Sta91, Mar92}. The mass loss is one of the important parameters, which affect the stellar evolution, especially for the low-mass stars like red dwarfs. In the case of magnetic active stars, the chromospheric activity of the stars are relatively very strong due to the their high rotational speed, when the stars have just came to the Zero Age Main Sequence (ZAMS), which led the rate of the mass loss to increase. The observations indicate that the mass loss rate of the Sun is about $2\times10^{-14}$ $M_{\odot}$ per year \citep{Ger05}, however it is about several $10^{-10}$ $M_{\odot}$ per year in the case of young UV Ceti type stars \citep{Sch00, Boy70}. This indicates that the UV Ceti type stars lose the largest part of their angular momentum in the beginning of the main-sequence life \citep{Mar92}.

However, we do not exactly know how the flare process causing mass loss with the high rate is working on a UV Ceti type star. The highest energy detected from two-ribbon flares that is the most powerful flares occurring on the sun is found to be $10^{30}$ - $10^{31}$ erg \citep{Ger05, Ben08}. Beside the Sun, RS VCn binaries, whose components are generally from the spectral types G or K giants, also exhibit the flare activity. Although their flare activity are generally observed in the radio or X-Ray bands \citep{Pan12}, there are several white-light observations of the visible region, in which the white-light flare light curves have been obtained \citep{Mat92}. Examining these white-light light curves, the energies emitting in the flares detected from RS CVn binaries are about $10^{31}$ erg \citep{Hai91}. Apart from RS CVn binaries, the ground-based observations demonstrated that the events are more frequently occurring on the surfaces of the dMe stars. In the case of dMe stars, the flare energy can reach a level from 10 times to 1000 times of the level reached in the solar case \citep {Gud09}. In fact the observations reveal that the flare energy level varies from $10^{28}$  erg to $10^{34}$  erg in the case of dMe stars \citep{Hai91, Ger05}. In addition, some dMe stars in the young clusters such as the Pleiades cluster and Orion association exhibit some powerful flare events, which energies reach $10^{36}$  erg \citep{Ger83}.

As it has been summarized above, there are so serious differences between the flare patterns, such as the flare energies and mass loss, of different type stars, such as the solar and dMe cases, that their flare process should be different from each other. However, the flare process working in UV Ceti stars has been explained by the Standard Solar Dynamo in spite of all. Especially, the source of the flare energy in the case of dMe stars is generally assumed the magnetic reconnection processes \citep{Ger05, Hud97}. On the other hand, there are lots of the points waiting to be explained. In this point, comparing the flares obtained from the different type stars, all the similarities and differences should be demonstrated. For example, first of all, the sources of the differences in some parameters like the flare energy spectra or flare frequency should be found. In this aim, it should be analysed whether the differences in the flare parameters from a star to the next are caused by the some physical parameters of the source stars, such as stellar mass, age, etc. In this point, there is also one more question that it should be analysed whether being a binary system causes some effects on the flare activity occurring on the components.

To understand the real reasons of the differences in the flare parameters from a star to the next, it should be demonstrated the nature of these stars with all their internal layers. There is a few indicators to reveal the internal layers of the stars. One of them is the stellar pulsation. Unfortunately the pulsations can not be observable for each star. On the other hand, the initial analysis and some studies in the literature, such as \citet{Uyt11}, indicated that KIC 2557430 is one of the candidates for the eclipsing binaries with pulsating component. Moreover, the pulsation behaviour seen in the stars is very important pattern to understand the stellar itself and its evolution. According to the observations lasting as long as several decades indicate that there are several type pulsating stars such as Cepheid, $\gamma$ Doradus, and $\delta$ Scuti type pulsating stars in the Instability Strip in the Hertzsprung-Russell diagram, especially on the main sequence. All these types are separated by their locations in the Instability Strip from each other. Analysing the pulsation frequencies, which is generally known as stellar seismology called asteroseismology, the physical processes behind both the pulsating nature and stellar interiors can be revealed. This is why the pulsating stars have an important role to understanding stellar evolution \citep{Cun07, Aer10}.

In addition, a pulsating star could be a component in an eclipsing binary system. In this case, this pulsating star has more important role to figure out the stellar evolution process. This is because, in some cases, the pulsation features cannot be enough to reveal the entire stellar interior and its physical parameters. However, analysing the light curves of the eclipsing binaries, lots of parameters such as mass ($M$), radius ($R$), and $log g$, can be easily obtained \citep{Wil71, Wil90}. In this point, considering the results obtained from both pulsating and eclipsing behaviour, the physical natures of components can be easily identified \citep{Lam06, Pig06}. As it is well known there are many pulsating single stars in the Instability Strip. However, the number of the pulsating stars being a component in an eclipsing binary is very small \citep{Kim10}.

In this study, we analysed the light variations of KIC 2557430, which is one of the first samples for its kind. One component of this system is seen as a pulsating star and another one is a chromospherically active star with strong flare activity. However, this chromospherically active component has some different physical properties from its analogue UV Ceti type stars due to being a component in a binary system. In this aim, first of all, the frequencies of the variations out-of-eclipses were analysed to figure out source of the variations. Then, we did complete light curve analyses of the system for the first time in the literature in order to find out the physical properties of the components, using the PHOEBE V.0.32 software \citep{Prs05}, whose method depends on the 2003 version of Wilson-Devinney Code \citep{Wil71, Wil90}. Finally, the flares occurring on the chromospherically active component were detected to model the magnetic activity nature of the system, comparing the active component with its analogue. Unfortunately, the physical parameters of internal layers for a star can be determined by analysing the multicolour light curves in the case of the photometric data \citep{Soy13}. In this point, the data using in this study were taken from the Kepler Mission Database, which are the monochromatic data \citep{Bor10, Koc10, Cal10}. Because of this, the pulsation modes were not determined from available data. In the second step of the study, we will try to obtain the multicolour photometric data, then we can complete the pulsation analysis. In this paper, we just presented the pulsation frequencies determined from the available monochromatic data.

KIC 2557430 is listed as an Algol type eclipsing binary with a brightness of $V=11^{m}.63$,  in the SIMBAD Database. Although the system was listed in the Tycho Inpute Catalogue by \citet{Egr92} for the first time, there are so few studies about this system in the literature. The system was listed as 2MASS J19072286+3748571 in the 2MASS All-Sky Survey Catalogue, in which its JHK brightness were given as $J=13^{m}.574$, $H=12^{m}.926$, $K=12^{m}.782$ \citep{Kha01}. In the literature, there are several approaches for the temperature of the system, though it has been observed very long time in the Kepler Mission \citep{Bor10, Koc10, Cal10}. Considering the data taken by the Kepler Satellite \citep{Sla11} computed the inclination ($i$) of the system as 77$^\circ$.17 and the color excess as $E(B-V)=0^{m}.077$, while the temperature of the system was found to be 6248 K with the temperature ratio of 0.951 for the components. Examining all the data in the literature of KIC 2557430, \citet{Pin12} found its metallicity as $[Fe/H]=-0.34$ dex, and stated that the temperature of the primary component has a value between 6547-6248 K. \citet{Hub14} found the temperature of the system between 6539-6531 K, and computed the mass and radius of the primary component as $M=$1.117 $M_{\odot}$  and $R=$1.524 $R_{\odot}$. Using some calibrations obtained from the data taken by the 2MASS All-Sky Survey and the Kepler Mission, \citet{Arm14}  found the temperature of the system as 6913 K. In the literature, there are so few studies, in which the variations out-of-eclipses were analysed. The period of the system was found to be 1.297743 day for the first time by \citet{Uyt11}, who stated that KIC 2557430 is an eclipsing binary with a $\gamma$ Doradus-type component. \citet{Cou14}  confirmed that the orbital period of the system is 1.297747 day. However, \citet{Bal15}, who revealed the flare activity detected from the system, stated that the rotational period is 2.02 day for KIC 2557430 apart from the orbital period.

\section{Data and Analyses}
The data analysed in this study were taken from the Kepler Mission Database \citep{Sla11, Mat12}. The Kepler Mission is a space mission in the aim of finding out exo-planet. More than 150.000 targets have been observed in this mission from 2009 \citep{Bor10, Koc10, Cal10}. The quality and sensitivity of Kepler observations have the highest one ever reached in the photometry \citep{Jen10a, Jen10b}. Lots of variable stars such as new eclipsing binaries or pulsating stars, etc. have been also discovered apart from the exo-planets in this mission \citep{Sla11, Mat12}. In addition, the observations indicate that an important part of single or double stars discovered among these newly discoveries, which some of them are the eclipsing binaries, exhibiting chromospheric activity \citep{Bal15}.

Considering the analyses of the flare activity in the study, the data were taken in short cadence format from the Database. All the available data reveal that KIC 2557430 has been observed in two observing parts. One of them was lasting about one month between HJD 24 55002.5109509 - 24 55033.3041338, while the second one was again lasting about one month but between HJD 24 55093.2155524 - 24 55123.5566582. Because of this, the analyses were sometimes done in two steps, and the first month observations were called as Part 1 data and the second month observations were called as Part 2 data in these analyses.

\begin{center}
\begin{equation}
\label{eq:four}
JD~(Hel.)~=~24~54954.461036\pm0.017907~+~1^{d}.12977364\pm0^{d}.0000015~\times~E
\end{equation}
\end{center}

The data were phased by using both the epoch and the orbital period given by Equation (1), which were taken from the Kepler Mission database and the light curves versus phase are shown in Figure 1. The entire light curves were shown in the bottom panel, while the light curves out-of-eclipses are shown in the upper panel for better visibility of light variations. As it is seen from the upper panel of the figure, the light curves out-of-eclipses are changing from one cycle to the next. Three variations are seen from the figure. The one of them is the primary and secondary minima due to the eclipses; the second one is a sinusoidal variation, and finally, the last one is instant-short term flare events. In the analyses, we arranged the data in suitable format considering the light curve analysis, sinusoidal variation analysis and the minimum time variation ($O-C$).

\subsection{Variability Out-Of-Eclipses: Pulsation}
To examine the sinusoidal variations out-of-eclipses, both all the minima due to the eclipses and all the flare events, sudden - rapid increasing in the light, were removed from the entire light curves. Thus, the remaining light curves were obtained, which is hereafter called as the pre-whitened light curves. For this purpose, the data of all primary minima observations between the phases of 0.955 - 0.045 and all secondary minima observations between 0.455 and 0.545 in phase were removed from the general light curve data. Comparing the consecutive light curve cycles in the pre-whitened light curves revealed that the consecutive cycles are absolutely different from each other. It is seen that the phases and levels of maxima and minima are rapidly changing from one cycle to the next.

In the first place, the source of these variations is seen as the rotational modulation due to the stellar cool spots. However, considering both the orbital period of 1.297747 day and the flare activity, there must be another source affected these variations out-of-eclipses. If the sinusoidal variations out-of-eclipses were caused due to just spot activity, the shape of the consecutive pre-whitened light curves should not been absolutely changed in 1.297747 day from one cycle to the next. Because of this, there must be one more source like stellar pulsation. In fact, \citet{Uyt11} indicated that KIC 2557430 is an eclipsing binary with a $\gamma$ Doradus-type component.

In this purpose, the pre-whitened light curve data were analyzed with the PERIOD04 Program \citep{Len05}, which is depends on the method of Discrete Fourier Transform (DFT) \citep{Sca82}. The results obtained from DFT were tested by two other methods. One of them is CLEANest, which is another Fourier method \citep{Fos95}, and the second method is the Phase Dispersion Minimization (PDM), which is a statistical method \citep{Ste78}.

\begin{center}
\begin{equation}
L(\theta)= A_{0} ~ + ~ \sum_{\mbox{\scriptsize\ i=1}}^N ~ A_{i} ~ cos(i \theta) ~ + ~ \sum_{\mbox{\scriptsize\ i=1}}^N ~ B_{i} ~ sin(i \theta)
\end{equation}
\end{center}
where $A_{0}$ is the zero point, $\theta$ is the phase, while $A_{i}$ and $B_{i}$ are the amplitude parameters.

Considering both the errors of parameters and also the signal to noise ratios, the results of the frequency analysis by the PERIOD04 Program \citep{Len05}, which is based on Equation (2) described by \citet{Sca82, Len05}, indicate that there are 50 different frequencies. The normalized power-spectrums distribution obtained from the Discrete Fourier Transform \citep{Sca82} is shown in Figure 2, while the obtained parameters are listed in Table 1. In the table, the frequency numbers are listed in the first column, while the obtained frequency values are listed in the second column. The amplitudes are listed in the fourth column, while the phase values are listed in the sixth column. The error of each parameter is listed in just next column in the table. In addition, the signal to noise ratio ($S/N$) of each frequency is listed in the last column.

Examining each frequency of all 50 frequencies, it is seen that the frequencies F1 and F4 are relevant to the orbital period of system. The frequency F1 is relevant to the orbital period itself, while the frequency F4 is relevant to the half of this period. In this case, these two frequencies must be relevant to a variation caused any rotational modulation possibly due to the stellar cool spots. Therefore, 48 frequencies must be relevant to the stellar pulsation apart from frequencies F1 and F4.

Using the obtained frequencies the synthetic light curve was derived for the variation out-of-eclipses by Equation (2) \citep{Sca82, Len05}. This synthetic curve and the pre-whitened are shown in Figure 3. As it is seen from the figure, the synthetic curve perfectly modelled the pre-whitened curve, which indicates that the analysis correctly worked.                 

\subsection{Variability Out-Of-Eclipses: Stellar Spot Activity}
After removing all the synthetic sinusoidal waves associated with just 48 frequencies of the pulsations from all the pre-whitened light curves, we got the residual variations, which are associated with the rest two frequencies listed as F1 and F4 in Table 1. However, as it was shown in Figure 4, these residual data show very striking variation because there is still a sinusoidal variation. However, this variation is not stable. As it is seen from figure 3, the shape of the residual light curve obtained with 48 frequencies is not changing along several cycles of orbital period. This is because this variation is caused due to stellar pulsation. On the other hand, it is clearly seen that the shape of the residual light curve obtained with the frequencies of F1 and F4 is absolutely changing from the Part 1 data to the Part 2 data. As it is seen from Figure 4, the phase of the residual variation and its amplitude are absolutely changed from the Part 1 data to the Part 2 data. It seems to be that there are two minima, which the deeper one is located in the later phases, in the residual light curve of Part 1 data, while the residual light curve of the Part 2 data has just one minimum, which is located in earlier phases. However, the latter light curve is seen more asymmetric. There are two months from the observations of Part 1 to those of Part 2. In this time interval, the phase of the deeper minimum has migrated from the phase of 0.64 to 0.24.

We wanted to check whether this shape change is a systematic slow variation or unsystematic sudden. For this purpose, the consecutive cycles of the residual light curve was examined for Part 1 and Part 2, separately. It was seen that the shape of the residual light curve is so slowly changing from one cycle to the next that the variation needs eight or nine cycles to be noticeable. Because of this, the residual data of each part observations were separated to three subsets that each of them contains about ten-day data. In the upper panel of Figure 4, the data of subsets were plotted in different color for both part observations. To better view, each of the ten-day data was averaged phase by phase with interval of 0.01. The averaged light curves were plotted in the bottom panel of Figure 4. As it is seen from this panel, the shape of the residual light curve is clearly changing from the one ten-day data to the next. In Figure 4, the filled black points represent the first ten-day data, while the filled blue points represent the second ten-day data, and finally, the filled red points represent the last ten-day data. As it is seen the figure, the minima of the residual light curves are migrating toward the earlier phases from the first ten-day data to the last ten-day data in both part observations. These small migrations in each part support that the shape of the residual light curve is slowly and systematically changing from the Part 1 data to the Part 2 data.

Apart from the pulsation, this variation can be caused by a third body or the stellar cool spot activity. In the first places, the reason of the variation cannot be third body. As it can be seen from the next section, although there are some light excess in the total light of the system due to a third body, but this is so small excess to show any sinusoidal variation seen in Figure 4. However, considering the existence of flare activity, it means that there are the spotted areas migrating toward the earlier longitudes on a component.

\subsection{Light Curve Analysis}
Examining the entire light curves of KIC 2557430 observing by Kepler Mission along two months cycle by cycle, it was seen that there are three different variations, such as eclipses, flare and sinusoidal variation out-of-eclipses, in the light curves. However, the frequency analyses indicated that the sinusoidal variation is not just caused by the stellar pulsations, but also a stellar spot activity on a component. Stellar spot activity can be easily modelled in the light curve analyses, but the instant-short term variation like flare activity or complex cyclic variation in the different phase like pulsation cannot model in the light curve analyses. Because of this, the variations caused by both the flare activity and pulsation waves from the all entire light curves were removed, before the analysing the light curve. The flare activity as an instant-short term is clear to easily detect with their distinct light variation. Therefore, we first of all removed all the variations due to the flare activity from the data. After then, the synthetic light curve, derived with 48 frequencies obtained from the frequency analyses, shown in Figure 3 was also removed from the all entire light curves. However, the residual sinusoidal variation caused by both the frequencies listed as F1 and F4 in Table 1 was leaved to be modelled in the light curve analysis. However the shape of the residual sinusoidal variation is different in the Part 1 and 2 observations. Because of this, in the light curve analysis, the residual sinusoidal variation was modelled separately for the Part 1 and Part 2 data.

Using the PHOEBE V.0.32 software \citep{Prs05}, which is employed in the 2003 version of the Wilson–Devinney Code \citep{Wil71, Wil90}, we analyzed the light curves obtained from the averages of all the detrended short cadence data. We attempted to analyse the light curves in various modes, including the detached system mode (Mod2), semi-detached system with the primary component filling its Roche-Lobe mode (Mod4), and semi-detached system with the secondary component filling its Roche-Lobe mode (Mod5). If the obtained stellar absolute parameters and the stellar evolution models are considered together in the analysis, the initial test demonstrated that an astrophysically reasonable solution was obtainable only in the detached system mode; no results that were statistically consistent with reasonable solutions could be obtained in any of the other modes.

Although there is no much more detailed studies depending on the light curve analysis for the system in the literature, lots of temperature values were given for the system, but there is no any clear vision about the temperature of the system. Because of this, taking each temperature value given in the literature as the temperature of the primary component respectively in the initial tests, we tried to find out which temperature value is the correct one. These tests indicated that an astrophysically reasonable solution can be obtainable only taking the temperature value of 6913 K given by \citet{Arm14} for the primary component. Thus, the temperature of the primary component was fixed to 6913 K, while the temperature of the secondary component was taken as adjustable parameter. Considering the spectral type corresponding to this temperature, the albedos ($A_{1}$ and $A_{2}$) and the gravity-darkening coefficients ($g_{1}$ and $g_{1}$) of the components were adopted for the stars with the convective envelopes \citep{Luc67, Ruc69}. The nonlinear limb-darkening coefficients ($x_{1}$ and $x_{2}$) of the components were taken from \citep{van93}. In the analyses, their dimensionless potentials ($\Omega_{1}$ and $\Omega_{2}$), the fractional luminosity ($L_{1}$) of the primary component, the inclination ($i$) of the system, the mass ratio of the system ($q$), and the semi-major axis ($a$) were taken as the adjustable free parameters. In addition, the fractional luminosity of the third body is also taken as the adjustable free parameter.

In addition, we modelled the residual sinusoidal variations out-of-eclipses by two cool spots on the secondary component in the PHOEBE V.0.32 software. Moreover, it was seen that the light curve analysis also gives a third light contribution of 0.339$\%$. All the parameters obtained from the light curve analysis were listed in Table 2, while the synthetic light curve derived with these parameters is shown in Figure 5. It must be noted that the temperature of the secondary component was found to be 6271$\pm$1 K. The error of the temperature is not realistic value, this is because the temperature error is statistically found in the process of the used code.

In the light curve analysis, the luminosity of the primary component was found to be 6.75791 $L_{\odot}$. Taking into count both its temperature and the luminosity, the primary component was plotted in the $log(T_{eff})$-$log(L/L_{\odot})$ plane in Figure 6. As it is seen from the figure, the primary component is located among the $\gamma$ Doradus type stars in the Instability Strip derived with the parameters taken from \citet{Gir00, Rol02}.

\subsection{Orbital Period Variation: $O-C$ Analysis}
Using the available short cadence detrended data of the system in the Kepler Mission Database \citep{Sla11, Mat12}, the minima times were computed without any extra correction on these detrended data. The minima times were computed with a script depending on Kwee $\&$ van Woerden Method described by \citet{Kwe56}, which considers just minima with its branches to compute the time of the minima, using a theoretical fit derived by the Least Squares Method. For all the minima times, the differences between observations and calculations were computed to determine the residuals $(O-C)_{I}$. Some minima times have very large error, for which the light curves were examined again. It was seen that there is a flare activity in these minima, then, these minima times were removed from the list. Finally, 89 minima times were obtained from the available short cadence detrended data. Using the regression calculations, a linear correction was applied to the differences, and the $(O-C)_{II}$ residuals were obtained. After the linear correction on $(O-C)_{I}$, new ephemerides were calculated as following:

\begin{center}
\begin{equation}
JD~(Hel.)~=~24~54954.46175(2)~+~1^{d}.1297728(3)~\times~E.
\end{equation}
\end{center}

All the calculated minima times, $(O-C)_{II}$ calculated from the differences between observations and calculations ($O-C$) were listed in Table 3. The minima times, epoch, minimum type and $(O-C)_{II}$ residuals are listed in the table, respectively. It is seen an interesting phenomenon in the variation of the $(O-C)_{II}$ residuals versus time. The $(O-C)_{II}$ residual variations are shown in Figure 7. A similar phenomenon has been recently demonstrated for chromospherically active other systems by \citet{Tra13, Bala15}.

\subsection{Flare Activity and the OPEA Model}
To understand the flare behaviour of the system, first of all, it needs to determine the flares from the available data. Then, it needs to determine and model the quiescent levels at the moment of the flares. For this purpose, all the primary minima between the phases of 0.955 - 0.045 and all the secondary minima between 0.455 and 0.545 in phase were removed from the entire light curves. Then, the observations with large error caused by some technical problems were also removed from the light curves.

To compute the parameters of a flare, it needs to determine where a flare is beginning and end. In this aim, we attempted to derive the quiescent levels in the light curves. In this point, the synthetic models obtained by the frequency analyses were used. All the sinusoidal variations, which are occurring due to the stellar pulsation and also the stellar spot activity, were derived for the light curves out-of-eclipses. In this point, the synthetic model was re-derived with the Fourier transform again, using all the found 50 frequencies for this time. In each point of the entire light curve, this synthetic model was assumed as the quiescent level of the light curve without any flare events. Some samples of the detected flares and the synthetic quiescent light curve at the moment are shown in Figure 8. The filled circles represent the observations, while the red lines represent the synthetic quiescent level of the light curve.

Using this synthetic model, the flare rise time ($T_{r}$), the decay time ($T_{d}$), amplitude of the flare maxima, and the flare equivalent duration ($P$) were computed for each flare, after defining both the flare beginning and the end for each flare. All these parameters are listed in Table 4.

Considering all the available short cadence data given in the Kepler Database, 69 flares were detected in total. In the analysis, the equivalent duration of each flare was computed using Equation (4) taken from \citet{Ger72}:

\begin{center}
\begin{equation}
P~=~\int[(I_{flare}-I_{0})/I_{0}]~dt
\end{equation}
\end{center}
where $P$ is the flare-equivalent duration in the observing band, while $I_{0}$  is the flux of the star in the observing band while in the quiet state. As it has just been described above, we computed the parameter $I_{0}$  using by the synthetic models derived with the Fourier transform. $I_{flare}$ is the intensity observed at the moment of the flare. Here it should be noted that the flare energies were not computed to be used in the following analyses due to the reasons described in detail by \citet{Dal10, Dal11}. Instead of the flare energy, flare equivalent duration has been used in the analysis. This is because of the luminosity term in the equation of flare energy, given by \citet{Ger72}. The luminosities of stars with different spectral types have large differences. Although the equivalent durations of two flares detected from two stars in different spectral types are the same, calculated energies of these flares are different due to different luminosities of these spectral types. Therefore, we could not use these flare energies in the same analysis. However, flare equivalent duration depends just on power of the flare. Moreover, the given distances of a star in different studies could be quite different. These differences cause the calculated luminosities become different.

In a result, obtained parameters, such as flare maximum times, equivalent durations, rise times, decay times and amplitudes of flare maxima, are listed from the first column to the last in Table 4, respectively.

Examining the relationships between the flare parameters, it was seen that the distributions of flare equivalent durations on the logarithmic scale versus flare total durations are varying in a rule. The distributions of flare equivalent durations on the logarithmic scale cannot be higher than a specific value for the star, and it is no matter how long the flare total duration is. Using the SPSS V17.0 \citep{Gre99} and GrahpPad Prism V5.02 \citep{Daw04} programs, \citet{Dal10, Dal11} demonstrated that the best function is the One Phase Exponential Association (hereafter OPEA) for the distributions of flare equivalent durations on the logarithmic scale versus flare total durations. The OPEA function \citep{Mot07, Spa87} has a $Plateau$ term, and this makes it a special function in the analyses. The OPEA function is defined by Equation (5):

\begin{center}
\begin{equation}
y~=~y_{0}~+~(Plateau~-~y_{0})~\times~(1~-~e^{-k~\times~x})
\end{equation}
\end{center}
where the parameter $y$ is the flare equivalent duration on a logarithmic scale, the parameter $x$ is the flare total duration as a variable parameter, according to the definition of \citep{Dal10}. In addition, the parameter $y_{0}$ is the flare-equivalent duration in on a logarithmic scale for the least total duration, which it means that the parameter $y_{0}$ is the least equivalent duration occurring in a flare for a star. Logically, the parameter $y_{0}$ does not depend on only flare mechanism occurring on the star, but also depends on the sensitivity of the optical system used for the observations. In his case, the optical system is optical systems of the Kepler Satellite. The parameter $Plateau$ value is upper limit for the flare equivalent duration on a logarithmic scale. \citet{Dal11} defined $Plateau$ value as a saturation level for a star in the observing band.

After the OPEA model was derived for all 69 flares detected from KIC 2557430, it was seen that the correlation coefficient squared ($R^{2}$) is very low, while the probability value ($p-value$) is found to be very high. It means that the model does not perfectly fit the distributions. In fact, it had been seen that the distributions of flare equivalent durations on the logarithmic scale ($log P$) versus flare total time ($T_{t}$) split into two groups in the $T_{t}-log(P)$ plane. Especially, this dissociation gets much clearer to be seen for the flares, whose total flare time is longer than 1400 s. Because of this, the flares with the total times longer than 1400 s were split into two groups. Then, using the least-squares method, the OPEA models were derived separately for two groups. In addition, the confidence intervals of 95$\%$ were also derived in these models. In the second step, considering the derived two OPEA models with the confidence intervals of 95$\%$, the flares with the total times shorter than 1400 s were separated into these two groups. As a result, all the distributions of flare equivalent durations on the logarithmic scale ($log P$) versus flare total time ($T_{t}$) and all the derived models with the confidence intervals of 95$\%$ are shown in Figure 9. In the figure, the filled red circles represent the flares called as Group 1 in this paper, while the filled blue circles represent the flares of Group 2. Using the least-squares method, the parameters of both models were computed and listed in Table 5. The $span$ value listed in the table is difference between $Plateau$ and $y_{0}$  values. The $half-life$ value is equal to $ln2/K$, where $K$ is a constant expressing in the same units as the $x$ value, at the model reaches the $Plateau$ value \citep{Daw04}. In other words, the $n \times half-life$ parameter is half of the minimum flare total time, which is enough to the maximum flare energy occurring in the flare mechanism.

The OPEA models derived for both groups was tested by using three different methods, such as the D'Agostino-Pearson normality test, the Shapiro-Wilk normality test and also the Kolmogorov-Smirnov test, given by \citet{Dag86} to understand whether there are any other functions to model the distributions of flare equivalent durations on the logarithmic scale versus flare total durations. In these tests, as it is listed in Table 5, the probability value called as $p-value$ was found to be $p-value<0.001$ and this means that there is no other function to model the distributions of flare equivalent durations \citep{Mot07, Spa87}. Therefore, as it is seen from the correlation coefficient squared ($R^{2}$) obtained for both models, the separation of the flares as two groups are statistically real.

In the Kepler Mission program, KIC 2557430 was observed along 61.134289 day (1467.22299 hour) in total, from HJD 24 55002.5109509 to 24 55033.3041338 and from 24 55093.2155524 to 55123.5566582. The significant 69 flares were detected in total from the available data. As it is listed in Table 6, 40 samples of all flares belong to Group 1, while the 29 of them belong to Group 2. \citet{Ish91} described two frequencies for the stellar flare activity. These frequencies are defined as given by Equations (6) and (7):

\begin{center}
\begin{equation}
N_{1}~=~\Sigma n_{f}~/~\Sigma T_{t}
\end{equation}
\end{center}

\begin{center}
\begin{equation}
N_{2}~=~\Sigma P~/~\Sigma T_{t}
\end{equation}
\end{center}
where $\Sigma n_{f}$ is the total flare number detected in the observations, and $\Sigma T_{t}$ is the total observing duration, while $\Sigma P$ is the total equivalent duration obtained from all the flares. In this study, both $N_{1}$ and $N_{2}$ flare frequencies were computed for all flares and were also computed separately for both groups. All the results of the flare frequencies are listed in Table 6.

\section{Results and Discussion}
In this study, we exerted remarkable effort to figure out the nature of the KIC 2557430, which is classified as Algol type binary in the SIMBAD Database, taking the observational data from the database of the Kepler Mission \citep{Sla11, Mat12}. The initial analyses indicate that there are three type variations, such as eclipses, sinusoidal variations and flare activity, different from each other. The data were analysed in the suitable ways to figure out these sources.

First of all, examining the variations out-of-eclipses indicates that both the shapes and the phases of the wave minima are changing from one cycle to the next, which cycles are computed with the orbital period. Considering the flare activity detected from the system, this variation seems to be caused by the stellar cool spots on the active component. However, the orbital period of the system is given as 1.297747 day in the literature \citep{Cou14}. In addition, the rotational period is given as 2.02 day in just one study for KIC 2557430 \citep{Bal15}. In this case, the changing shape of the light curve out-of-eclipses cannot be explained by the stellar spot activity. There must be some other sources to rapid changes in the light curve out-of-eclipses. Because, it should not be expected that the configuration of the spotted areas on a star is changing to cause the radical changes in the light curve shape in the short time intervals such as 1.297747 or 2.02 day \citep{Ger05}. As it is known that one of the components is a $\gamma$ Doradus-type star \citep{Uyt11}. Considering the available information about the system, we analysed the light variation out-of-eclipses to find its characteristic frequencies. In total, 50 frequencies with the signal to noise ratio of 5.0 or with larger ratio than 5.0 were found in the analysis of the PERIOD04 program. Two of the found frequencies listed in Table 1, F1 and F4, are relevant to the orbital period of the system given by \citet{Cou14} and its half. Thus, the rest of them are relevant to the stellar pulsation. The later analyses indicated that there is no more astrophysically acceptable frequency. As it seen from Figure 3; the synthetic curve obtained from the found frequencies is really well fitted the observations.

Although the frequencies F1 and F4 are relevant to the orbital period, if the synthetic curve is obtained with the rest 48 frequencies, it is seen that this synthetic curve cannot perfectly fit the observations. If this synthetic curve is extracted from the observations, it is seen that a residual sinusoidal variation still remain in the light curve. It means that there is another source for the variation out-of-eclipses apart from the stellar pulsation. In addition, the analysis of this residual sinusoidal variation indicated an interesting result that the shape of this variation is dramatically changing from the Part 1 observations to the Part 2. Considering that these two frequencies are relevant to the orbital period, this residual sinusoidal variation out-of-eclipses must be caused by the rotational modulation due to the stellar cool spots probably. \citet{Kro52} demonstrated that the UV Ceti type stars also exhibit the stellar cool pot activity, too. This phenomenon called BY Dra syndrome in later years has been studied for several decades \citep{Bop73, Kun75, Vog75}. As it is in the case of Solar, it has been observed in lots of cases that a spotted area exhibits a cyclic migration on the surface of the star, while its location and size are varying at the moment, evolving on the surface by the time \citep{Fek02, Ber06, Ola06, Kor05, Ger05}. In the case of KIC 2557430, the residual sinusoidal variation must be caused by the spotted area, which is evolving while it is also migrating on the surface. This is explained why the shapes of the residual sinusoidal variation are different in Part 1 and 2. Consequently, considering the available flare observations, the sinusoidal variation found as the residual of the stellar pulsation in the pre-whitened light curves must be caused the rotational modulation due to the stellar chromospheric activity.

Although there are several approaches about the physical parameters of the system found using some calibration with the available data in the literature, there is no any entire light curve analysis for KIC 2557430. For this reason, the entire light curve of the system was analysed for the first time in the literature, using the PHOEBE V.0.32 software \citep{Prs05}, which uses the 2003 version of the Wilson–Devinney Code \citep{Wil71, Wil90}.

Although the temperatures given in the literature for the system vary from 5544 K \citet{Amm06} to 6913 K \citep{Arm14}. Trying each temperature value given in the earlier studies, we found that an astrophysically reasonable solution was obtainable only taking the temperature value of 6913 K for the primary component. Fixing this temperature for the primary component, the temperature of the secondary component was found to be 6271$\pm$1 K. The mass ratio ($q$) of the system was found to be 0.868$\pm$0.002, while the inclination ($i$) of the system was computed as 69.75$\pm$0.01. The dimensionless potentials ($\Omega_{1}$ and $\Omega_{2}$) of the components were found to be 5.8362$\pm$0.0010 and 5.0301$\pm$0.0009, while the fractional radii of the components were calculated as 0.2029$\pm$0.0004 and 0.2218$\pm$0.0006. On the other hand, the light curve analysis indicated the existence of the luminosity of 0.0034 $L_{\odot}$  for a third body.

According to these results, we revealed which component is located in the Instability Strip. In Figure 6, we plotted ZAMS and TAMS taken from \citet{Gir00} and the borders of the Instability Strip computed from \citet{Rol02} in the $log(T_{eff})$-$log(L/L_{\odot})$ plane. Apart from this, we also plotted some pulsating stars, which are the components in the binary systems, from two different types. Some of them are some $\delta$ Scuti stars taken from \citet{Soy06}, while some of them are $\gamma$ Doradus type stars taken from \citet{Hen05}. When the components of KIC 2557430 were plotted in this $log(T_{eff})$-$log(L/L_{\odot})$ plane, it was seen that the primary component is located among $\gamma$ Doradus type stars. However, according to the light curve analysis, the temperature of the secondary component was computed as $log(T_{eff})=$3.797 and its luminosity was calculated as $log(L/L_{\odot})=$0.729. In this case, the secondary component is located absolutely out of the Instability Strip in this plane. Therefore, as it was earlier stated by \citet{Uyt11}, one of components, is located in the Instability Strip and seems to be a $\gamma$ Doradus type star.

The minima times were computed for both the primary and secondary minima from the available short cadence data in the Kepler Database. After determining the differences $(O-C)_{I}$ between observations and calculations, the linear correction was applied on $(O-C)_{I}$, and $(O-C)_{II}$ residuals were obtained. In this point, the variation of $(O-C)_{II}$ residuals versus time exhibits two characteristic behaviours. First of all, as it is seen from the upper panel of Figure 7, the least-squares method indicated that all the $(O-C)_{II}$ residuals computed for both the primary and secondary minima show an inverse parabolic variation, which means normally that there is a mass transfer from the primary component to the secondary \citet{Pri85}. however, in the case of this system, the secondary component is filling its Roche Lobe, not the primary. Possible interpretation of the situation is that, considering existence of the flare activity exhibiting by KIC 2557430, the inverse parabolic variation must be caused due to the mass loss from the system. Secondly, if the variations of $(O-C)_{II}$ residuals obtained from the primary minima versus time and from the secondary minima are examined separately, it is clearly seen that the $(O-C)_{II}$ residuals of the primary and secondary minima vary in the same way, but in the different phases, asynchronously. This behaviour was demonstrated by \citet{Tra13} and \citet{Bala15}. According to their results, the synchronous variation but in opposite directions of the $(O-C)_{II}$ residuals is explained by the stellar spot activity on the component. In fact, the light curve analysis revealed the existence of two cool spotted areas on the secondary component.

The relations between the parameters indicate that the distributions of flare equivalent durations on the logarithmic scale versus flare total durations are varying in a rule. The distributions of flare equivalent durations had been modelled with the OPEA function by \citet{Dal11}. The authors also demonstrated that the OPEA models get a form depending on the ($B-V$) color indexes of the stars. According to their results, the OPEA models have the saturation levels, which are defined depending on the stellar ($B-V$) color indexes. In this study, available short cadence data given in the Kepler Database indicate that KIC 2557430 was observed along 61.134289 day (1467.22299 hour) in total. In total, 69 flares were detected from these data, and their parameters were computed. we initially tried to derive the OPEA model using all the detected 69 flares. However, the correlation coefficient squared ($R^{2}$) obtained from this initial OPEA model is very low, while its probability value ($p-value$) is very high. In this case, the model cannot be statistically acceptable. Examination of the distributions of flare equivalent durations revealed that the dissociation gets start in the data around the flare total time of 1400 s. Because of this, we split the data into two groups, depending on the flares with larger flare total time than 1400 s. Then, two OPEA models were derived for both flare groups. When the parameters derived from the OPEA models were statistically compared, these statistical comparisons indicated that these two OPEA models are absolutely different from each other. For example, the $Plateau$ value, which was defined as a saturation level for the model by \citet{Dal11}, was found to be 1.4336$\pm$0.1104 s for the flares of Group 1, while it was found to be 0.7550$\pm$0.0677 s for the flares of Group 2. Considering the errors of the $Plateau$ values, it is clearly seen that these $Plateau$ values cannot be equal to each other, which is very interesting because there is not any other sample for this case. According to the results obtained by \citet{Dal11, Dal12}, the flares detected from one star are always modelled with just one OPEA model. However, the flares detected from KIC 2557430 must be modelled with two different OPEA models. \citet{Dal11, Dal12} demonstrated that the OPEA models were formed depending on the stellar ($B-V$) color indexes. Thus, two OPEA models derived for the flares detected from KIC 2557430 definitely indicate existence of two flare sources in this system. In brief, the flares of Group 1 are come from one star, while the flares of Group 2 are come from another star in the system. If the $Plateau$ values of the models are considered, it will be understood that the flares of different groups must be come from two stars with different ($B-V$) color indexes. As a result, two components of KIC 2557430 must be separately flare stars.

On the other hand, here is a bit problematic case. The light curve analysis gave an astrophysically reasonable solution, if the temperature of the primary component was taken as 6913 K. In this case, the temperature of the secondary component is found to be 6271 K. Using the calibration given by \citet{Tok00}, the color index of the primary component was computed as $B-V=0^{m}.372$, it was found to be $B-V=0^{m}.516$ for the secondary component. In the study of \citet{Yol16a, Yol16b}, the $Plateau$ value of FL Lyr, whose ($B-V$) color index is $0^{m}.74$, was found to be 1.232 s, while it was computed as 1.951 s for KIC 9761199, whose ($B-V$) color index is $1^{m}.303$. In addition, the $Plateau$ value of KIC 7885570, whose ($B-V$) color index is $0^{m}.643$, was found to be 1.9815 s by \citet{Kun16},. Moreover, the $Plateau$ value was obtained as 3.014 s for EV Lac ($B-V=1^{m}.554$), 2.935 s for EQ Peg ($B-V=1^{m}.574$), and also 2.637 s for V1005 Ori ($B-V=1^{m}.307$) by \citet{Dal11, Dal12}. According to the results given in this brief summary, in the case of KIC 2557430, the flares of Group 1 seem to be come from the secondary component, considering its ($B-V$) color index with the $Plateau$ value. Thus, the flares of Group 2 should be come from the primary component. However, it is not possible due to its ($B-V$) color index according to the results obtained by \citet{Dal11, Dal12}. In this point, our thought is that the flares of Group 2 could be come from the third body the light curve analysis indicated. However, there is no data about what its ($B-V$) color index is. In the future, KIC 2557430 needs a spectral observation with high resolution.

From the analyses of the OPEA models derived for KIC 2557430, the $half-life$ values were found to be 2278.1 s for the flares of Group 1 and 1811.2 s for the flare of Group 2. The $half-life$ values found for KIC 2557430 are remarkably higher than those found for the UV Ceti type single stars. According to \citet{Dal11, Dal12}, it is 433.10 s for DO Cep ($B-V=1^{m}.604$), 334.30 s for EQ Peg and 226.30 s for V1005 Ori. As it is seen from this brief summary, the flares can reach the maximum energy level at their $Plateau$ value, when their total durations reach about $n \times 5$ minutes for a UV Ceti type single star. In the case of KIC 2557430, it needs $n \times 38$ minutes for the flares of Group 1, while it is $n \times 30$ minutes for the flares of Group 2. Similarly, in the cases of FL Lyr and KIC 9761199, it needs $n \times 39$ and $n \times 17$ minutes for these systems \citet{Yol16a, Yol16b}. In addition, it was found to be $n \times 66$ minutes for KIC 7885570 by \citet{Kun16}. In a result, KIC 2557430 is similar to FL Lyr in the point of the $half-life$ values.

The maximum flare total time ($T_{t}$) was found to be 9827.78 s for the flares of Group 1, while it was 9827.92 s for the flares of Group 2. In addition, the maximum flare rise time ($T_{r}$) was found to be 941.53 s for the flare Group 1, while it was 3059.99 s for the others. \citet{Yol16a, Yol16b} found that the maximum flare rise time ($T_{r}$) is 5179 s for the flares of FL Lyr, while it is 1118.1 s for the flares of KIC 9761199. Moreover, the maximum flare total time ($T_{t}$) was found to be 12770.62 s for the flares of FL Lyr, while it was 6767.72 s for the flares of KIC 9761199. At the moment, \citet{Kun16} found that the maximum flare rise time ($T_{r}$) is 7768.210 s and also the maximum flare total time ($T_{t}$) is 16890.30 s for KIC 7885570. On the other hand, for the single UV Ceti type stars \citet{Dal11}, found that the maximum flare rise time ($T_{r}$) is 2062 s for V1005 Ori, 1967 s for CR Dra. Similarly, the maximum flare total time ($T_{t}$) was found to be 5236 s for V1005 Ori and 4955 s for CR Dra. As it is seen from these results, the flare time scales indicate that the Group 2 flares detected from KIC 2557430 have the same character with KIC 9761199. However, the flares of Group 1 have the same character with the single UV Ceti type stars.

The Short Cadence data in the Kepler Mission Database reveal that KIC 2557430 was observed 1467.22299 hour. In these observations, 69 flares were detected. 40 of them was classified as Group 1, the rest of them was classified as Group 2. The flare frequencies of KIC 2557430 were computed as $N_{1}=$0.02726 $h^{-1}$ and $N_{2}=$0.00002 for Group 1, while they were calculated as $N_{1}=$0.01977 $h^{-1}$ and $N_{2}=$0.00001 for Group 2. Comparing these frequencies with those computed from single UV Cety type stars, it is seen that the flare energy level found for KIC 2557430 is remarkably lower than those found from them. For instance, the observed flare number per hour for UV Ceti type single stars was found to be $N_{1}=$1.331 $h^{-1}$ in the case of AD Leo, while it was found to be $N_{1}=$1.056 $h^{-1}$ for EV Lac. Moreover, $N_{2}$ frequency was found to be 0.088 for EQ Peg, while it was found to be $N_{2}=$0.086 for AD Leo \citep{Dal11}. However, according to \citet{Yol16a, Yol16b}, the flare frequencies were found as $N_{1}=$0.4163 $h^{-1}$ and $N_{2}=$0.0003 for FL Lyr ($B-V=0^{m}.74$), and $N_{1}=$0.0165 $h^{-1}$ and $N_{2}=$0.00001 for KIC 9761199 ($B-V=1^{m}.303$). In addition, $N_{1}$ and $N_{2}$ were computed as $N_{1}=$0.00362 $h^{-1}$ and $N_{2}=$0.00001 for KIC 7885570 ($B-V=0^{m}.643$) by \citet{Kun16}. It is clearly seen that the flare frequencies of KIC 2557430 for both groups are similar to the frequencies of KIC 9761199.

As a result, the frequency analyses and the light curve analysis indicate that the primary component of KIC 2557430 is most probably a $\gamma$ Doradus-type pulsating star. However, they also indicate that the secondary component of the system exhibits the cool spot activity on its surface. The analysis of the $(O-C)_{II}$ residuals reveals a mass loss from the whole system. In fact, there is a distinctive flare activity detected from the system. The analyses of the flare activity indicate the existence of two possible OPEA models, which means that the flares detected from KIC 2557430 are come from two different targets with different ($B-V$) color indexes. On the other hand, the primary component of the system is a $\gamma$ Doradus-type pulsating star. This component can certainly exhibit some flares, but it is too hot to exhibit often any huge flares like a UV Ceti star \citet{Ger05}. However, the light curve analysis indicated that the second component exhibits the spot activity, so it is possible that the secondary component can also exhibit the flare activity. In this point, there should be one more target to exhibit the flare activity.

In our opinion, this second target for the flare activity is the third body, whose existence was found in the light curve analysis. On the other hand, there is a handicap in this case. If the third bode found in the light curve analysis is able to exhibits some flare activity, this source must be an M dwarf at least. In this case, it is expected that the third light contribution in the total light must be much more than 0.339$\%$. However, the third body may be not a component in the system. This source can be a chromospherically active star located in the same direction with KIC 2557430 in the sky, but not in the same distance. If the case is real, the third body must be far away from the KIC 2557430 in the space. However, the stellar spot activity indicated by the residual frequencies of F1 and F4 should not be relate to the third body, because the F1 and F4 frequencies are correlated with the orbital period of the system.

It is seen that the problem about the third body and the second source of the flare activity needs the spectral observations with high resolution in the future. For this reason, KIC 2557430 should be observed spectroscopically with high time resolution in order to check the existence of the spectral lines for the third body.

\section*{Acknowledgments} The authors acknowledge generous allotments of observing time at Ege University Observatory and TUBITAK National Observatory of Turkey. We also wish to thank the Turkish Scientific and Technical Research Council for supporting this work through grant No. 116F349. We also thank the referee for useful comments that have contributed to the improvement of the paper.

\clearpage

\begin{figure*}[h]
\begin{center}
\includegraphics[scale=0.60, angle=0]{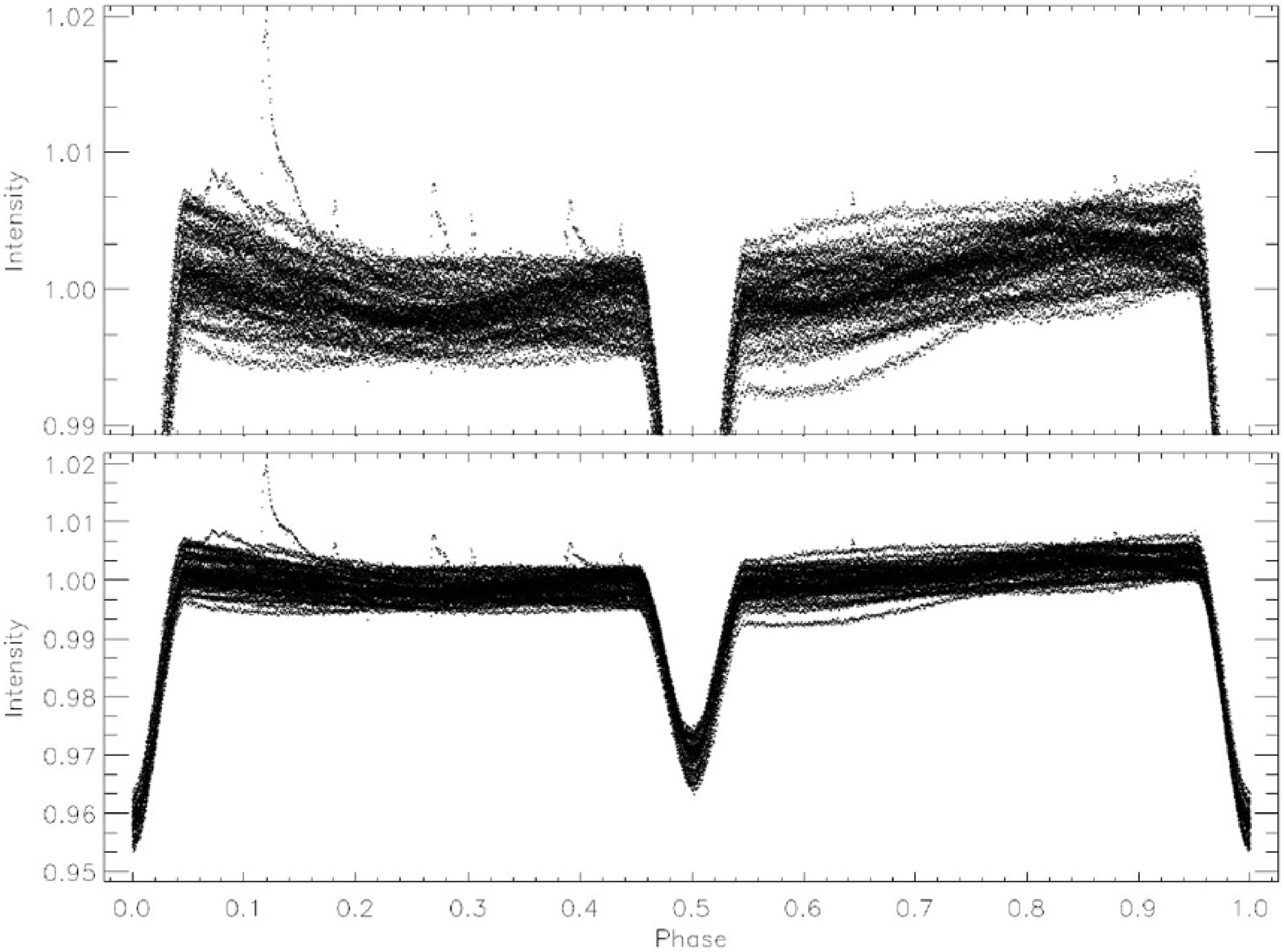}
\vspace{0.4 cm}
\caption{All the light curves of KIC 2557430 obtained from the available short cadence data in the Kepler Mission Database. The full of the light curves are shown in the bottom panel, while the maxima of the curves are shown in the upper panel to reveal the variations out of eclipses.}
\label{Fig. 1.}
\end{center}
\end{figure*}

\begin{figure*}[h]
\begin{center}
\includegraphics[scale=0.80, angle=0]{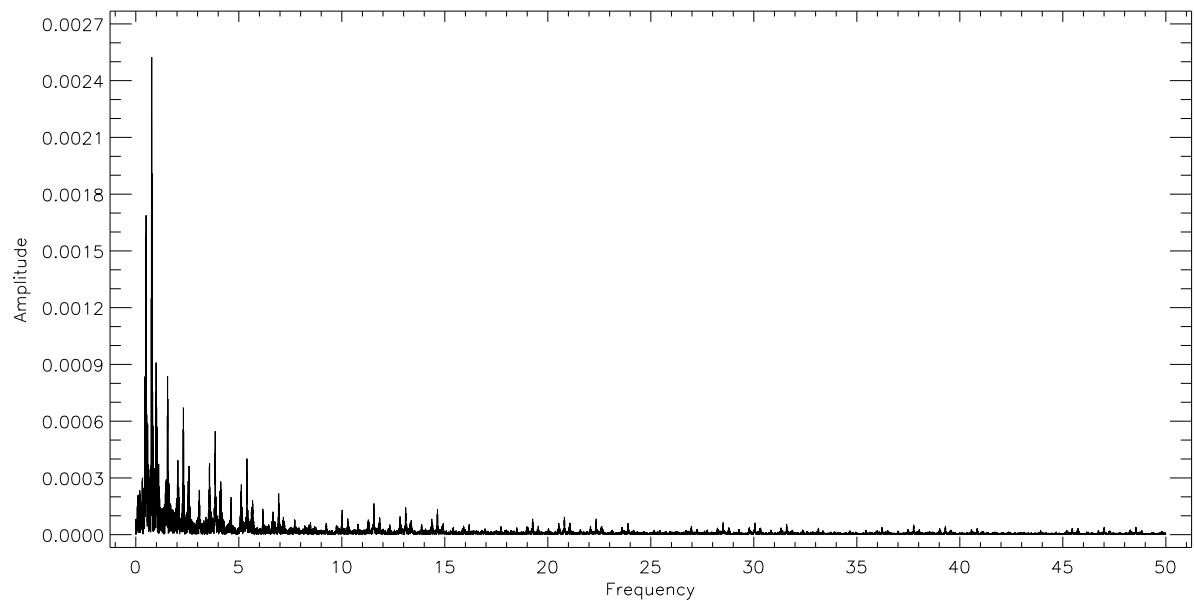}
\vspace{0.4 cm}
\caption{The normalized power-spectrums distribution obtained from the Discrete Fourier Transform \citep{Sca82}.}
\label{Fig. 2.}
\end{center}
\end{figure*}

\begin{figure*}[h]
\begin{center}
\includegraphics[scale=0.5, angle=0]{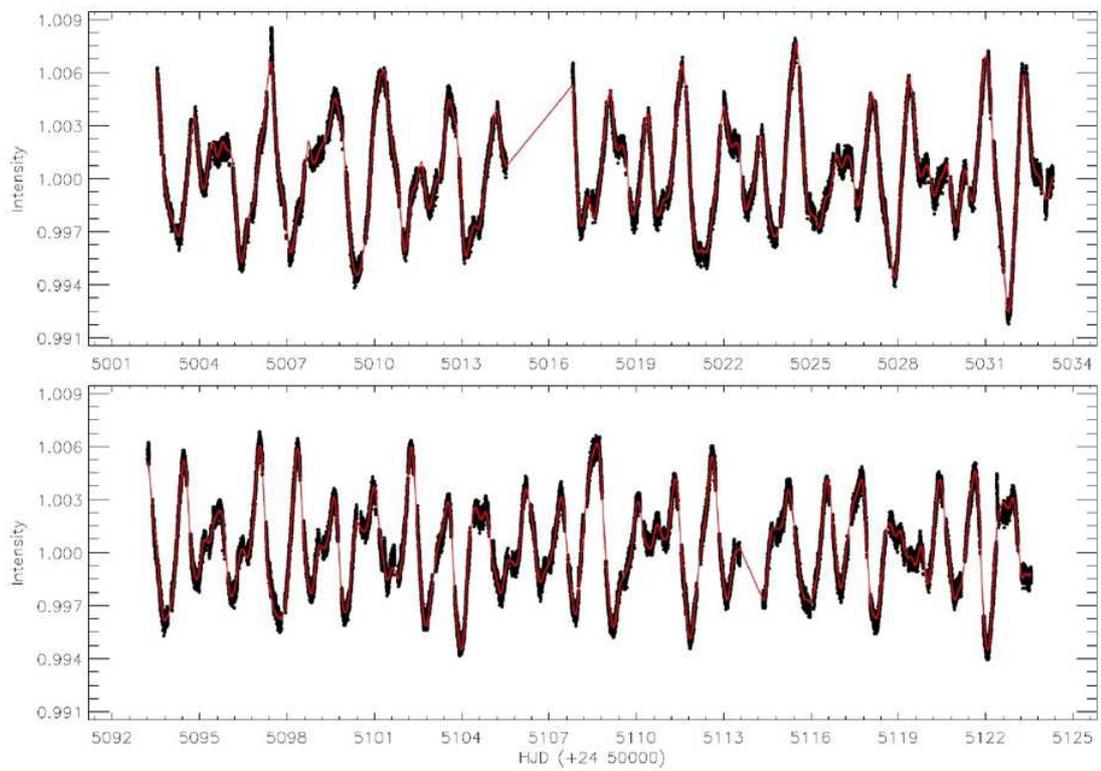}
\vspace{0.4 cm}
\caption{The variation out-of-eclipses and the synthetic model derived by the frequencies obtained from the Discrete Fourier Transform \citep{Sca82}. The filled circles represent the observations, while the red line represents the model.}
\label{Fig. 3.}
\end{center}
\end{figure*}

\begin{figure*}[h]
\begin{center}
\includegraphics[scale=0.5, angle=0]{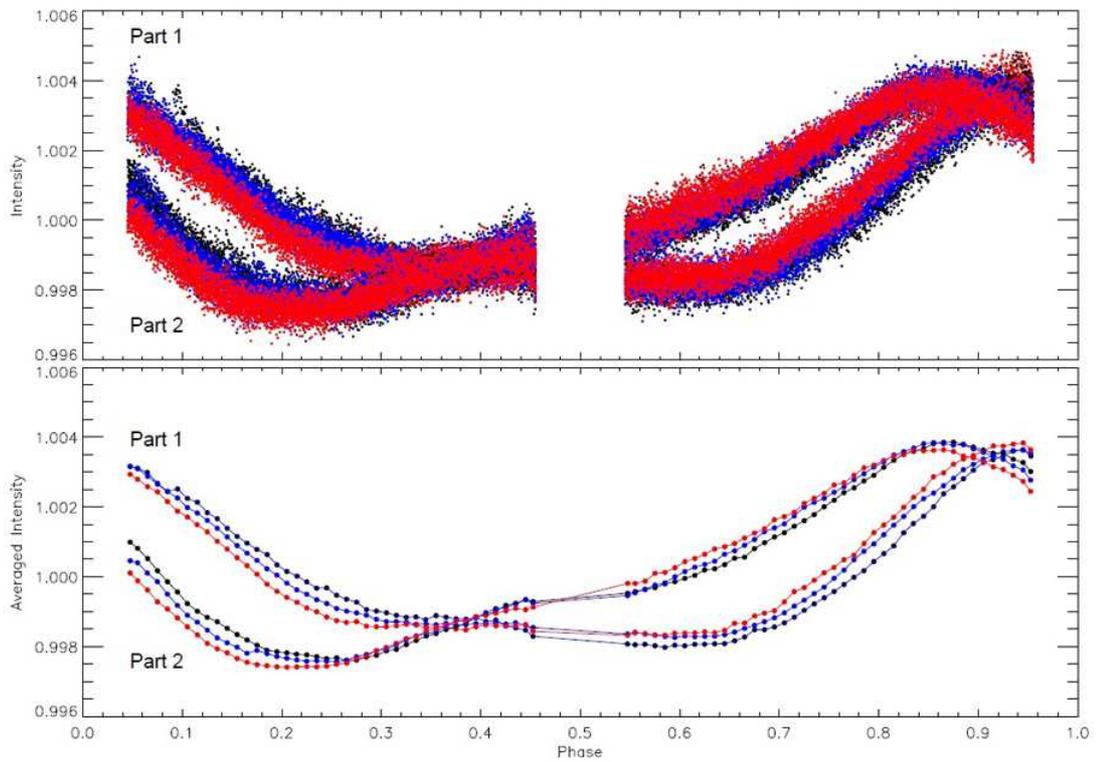}
\vspace{0.4 cm}
\caption{The residual variations obtained removing all the pulsation variations from all the pre-whitened light curves of KIC 2557430. In each panel, the black circles represent the first 10 days-observations, the blue circles represent the second 10 days-observations, while the red circles represent the third 10 days-observations for both Part 1 data taken between HJD  24 55002.51095 - 24 55033.30413 and Part 2 data taken between HJD 24 55093.21555 - 24 55123.55666.}
\label{Fig. 4.}
\end{center}
\end{figure*}

\begin{figure*}[h]
\begin{center}
\includegraphics[scale=0.5, angle=0]{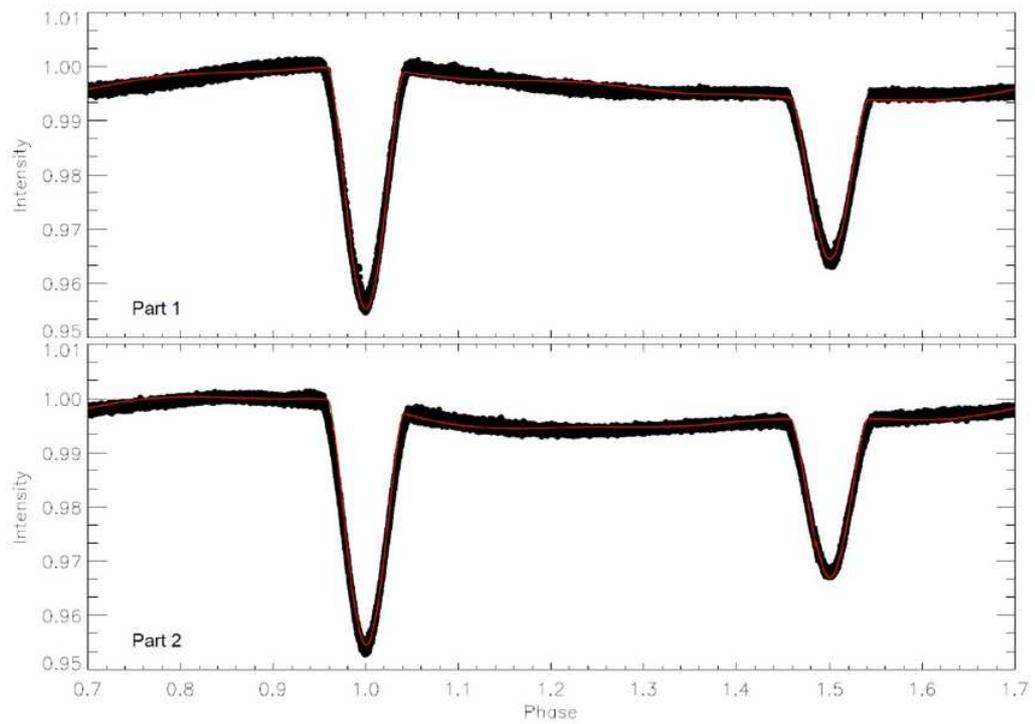}
\vspace{0.2 cm}
\caption{The observational and the synthetic light curves obtained from the light curve analyses of KIC 2557430 for the observations of Part 1 data taken between HJD  24 55002.51095 - 24 55033.30413 (upper panel) and Part 2 data taken between HJD 24 55093.21555 - 24 55123.55666 (bottom panel). The filled circles represent the observations, while the red line represents the model.}
\label{Fig. 5.}
\end{center}
\end{figure*}

\begin{figure*}[h]
\begin{center}
\includegraphics[scale=1.15, angle=0]{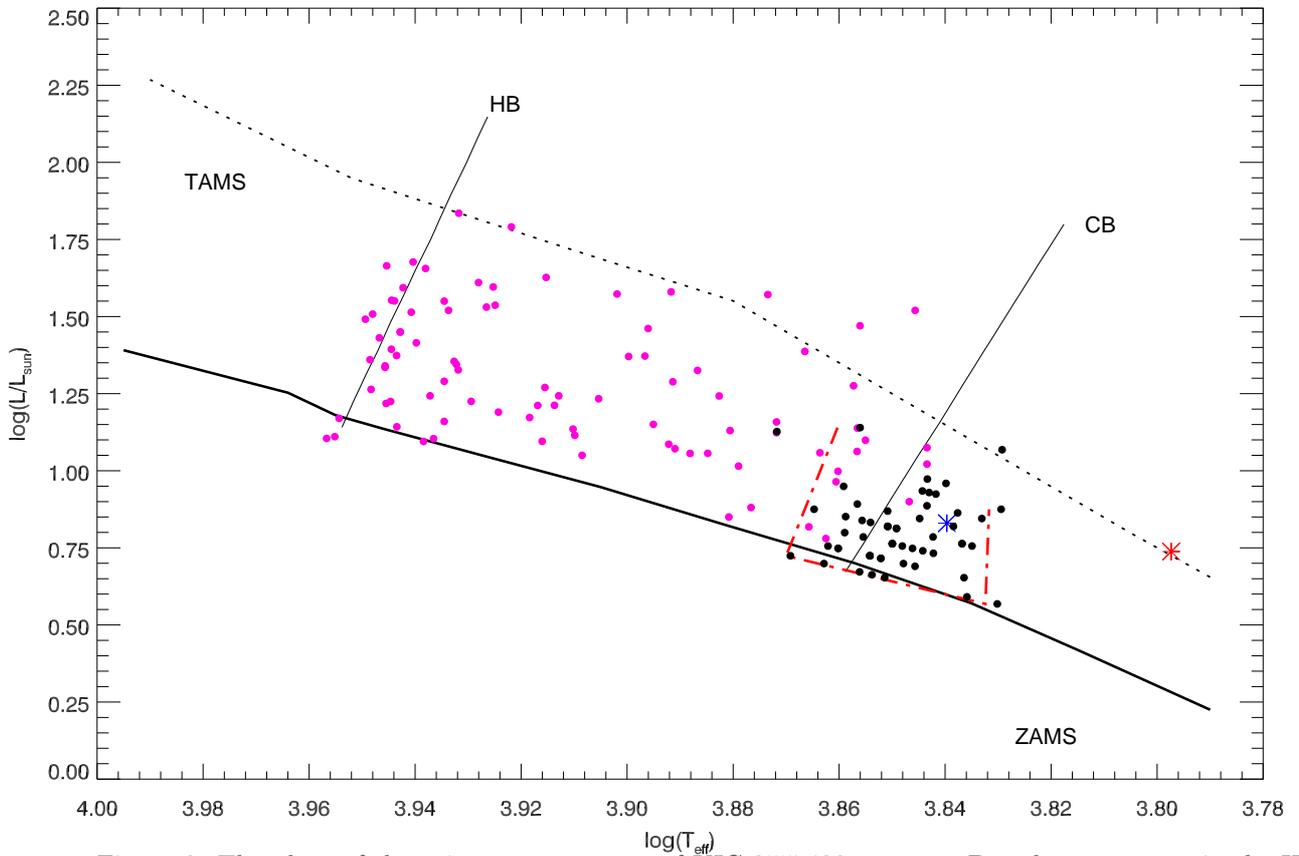}
\vspace{0.1 cm}
\caption{The place of the primary component of KIC 2557430 among $\gamma$ Doradus type stars in the HR diagram. In the figure, the small filled black circles represent $\gamma$ Doradus type stars listed in \citep{Hen05}. The asterisk represents the primary component of the system. The dash dotted lines (red) represent the borders of the area, in which $\gamma$ Doradus type stars take place. In addition we plotted the hot (HB) and cold (CB) borders of the $\delta$ Scuti stars for comparison. In the figure, the small filled pink circles represent some semi- and un-detached binaries taken from \citet{Soy06} and references there in. The ZAMS and TAMS were taken from \citet{Gir00}, while the borders of the Instability Strip were computed from \citep{Rol02}.}
\label{Fig. 6.}
\end{center}
\end{figure*}

\begin{figure*}[h]
\begin{center}
\includegraphics[scale=0.83, angle=0]{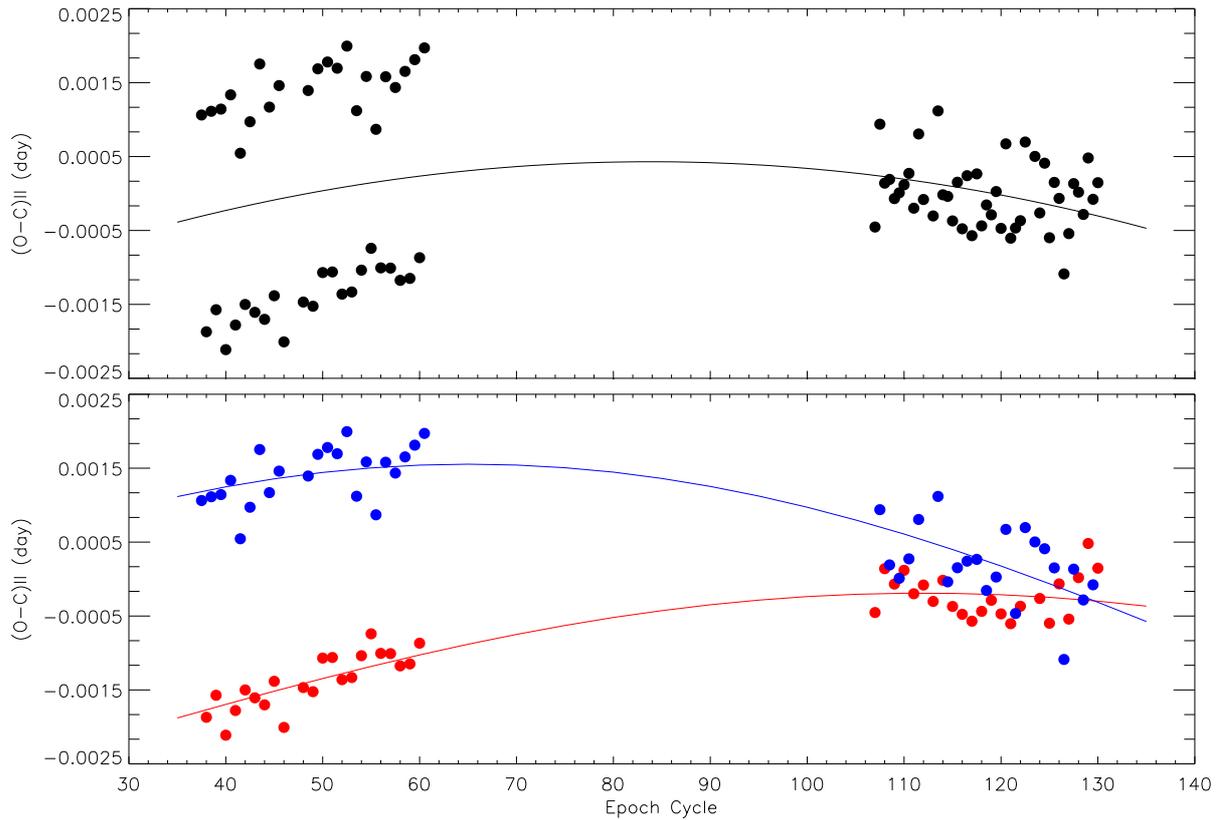}
\vspace{0.1 cm}
\caption{The variations of the minima times computed from the available short cadence data in the Kepler Database. All the $(O-C)_{II}$ residuals for both the primary and the secondary minima and the parabola fit derived from all the minima are shown in the upper panel, while the $(O-C)_{II}$ residuals for both the primary and the secondary minima are shown separately in the bottom panel. In both panels, the filled circles represent the $(O-C)_{II}$ residuals, while the lines represent the parabola fits. In the bottom panel, the red circles represent the secondary minima, while the blue circles represents the primary minima.}
\label{Fig. 7.}
\end{center}
\end{figure*}

\begin{figure*}[h]
\begin{center}
\includegraphics[scale=0.83, angle=0]{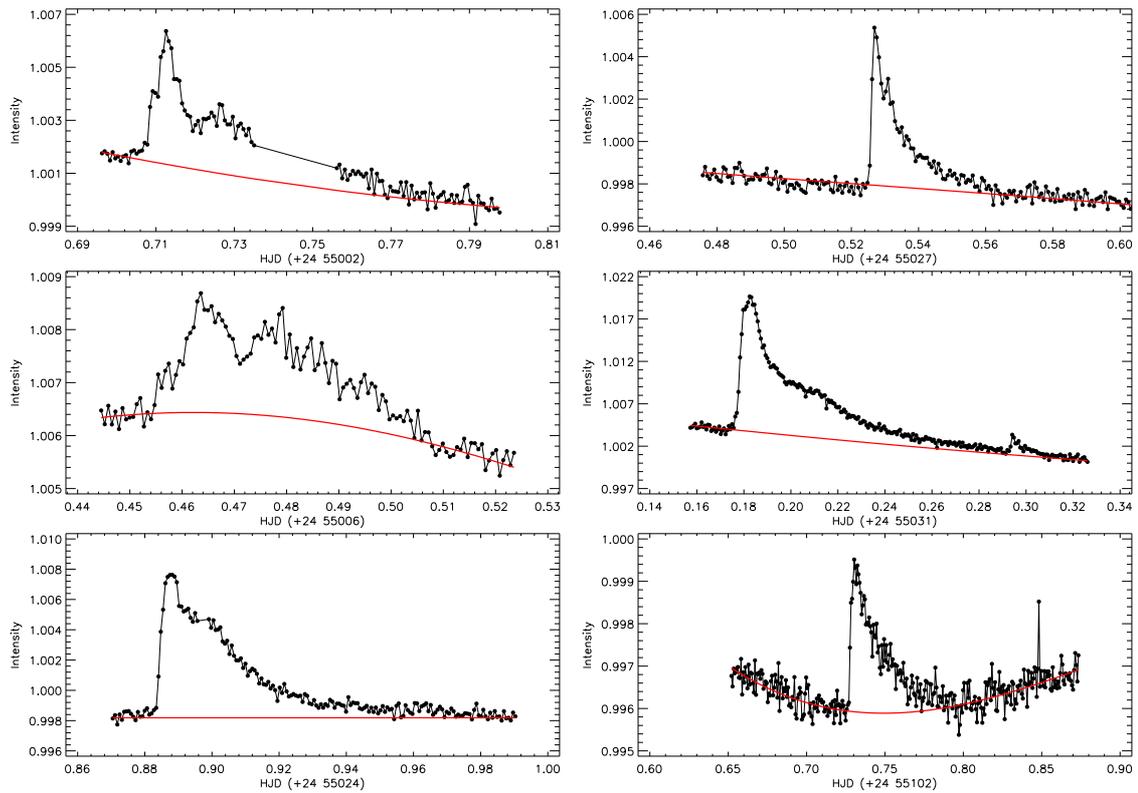}
\vspace{0.1 cm}
\caption{The flare light curve samples chosen from different parts of the short cadence data in the Kepler Database. In the figures, the filled circles represent the observations, while the red lines represent the synthetic curves assumed as the quiescent state of the star.}
\label{Fig. 8.}
\end{center}
\end{figure*}

\begin{figure*}[h]
\begin{center}
\includegraphics[scale=0.83, angle=0]{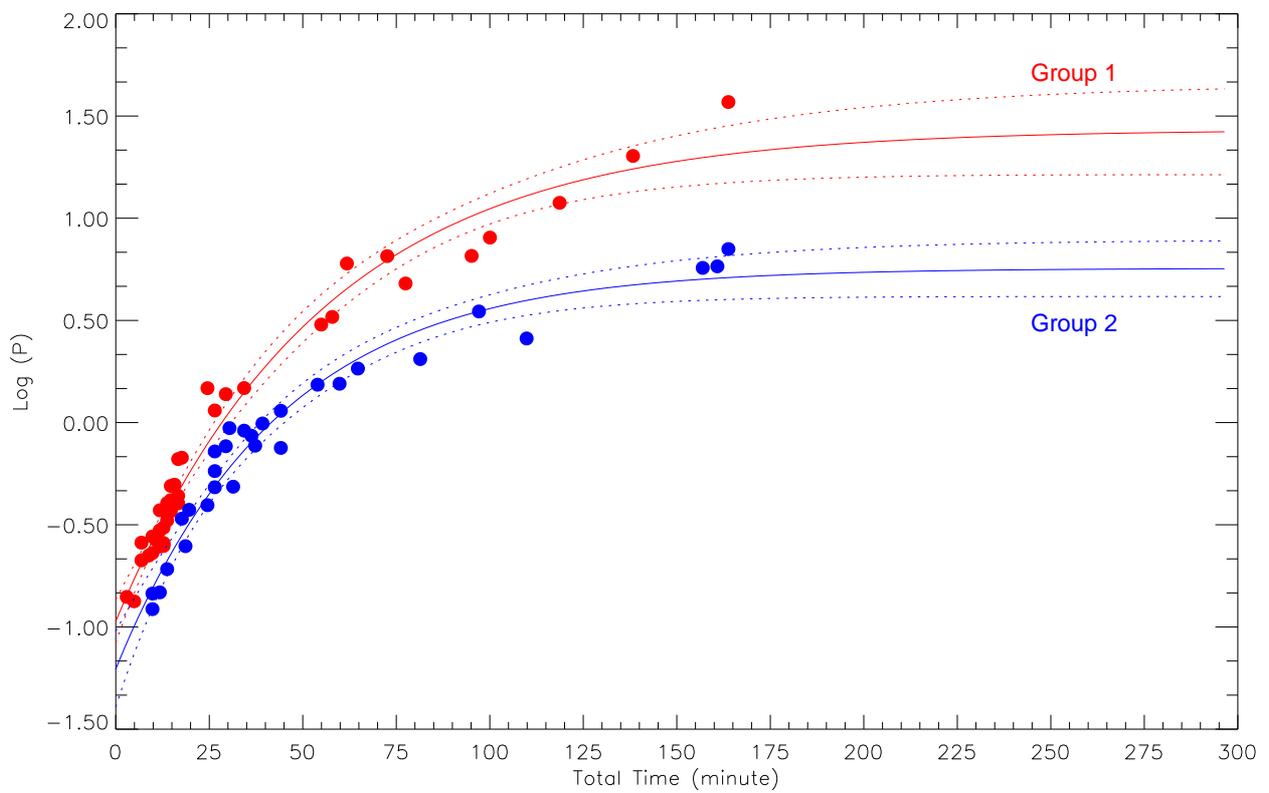}
\vspace{0.1 cm}
\caption{Using the least-squares method, the OPEA models derived from the detected 69 flares. In the figure, the filled circles represent the observations, while the lines represent the models.}
\label{Fig. 9.}
\end{center}
\end{figure*}

\clearpage

\begin{table*}
\begin{center}
\caption{All the frequencies and their parameters obtained from the Discrete Fourier Transform \citet{Sca82}, using the PERIOD04 Program.}
\begin{tabular}{@{}lccccccc@{}}
\hline
Nr.	&	Frequency	&	Fre. Error	&	Amplitude	&	Amp. Error	&	Phase	&	Pha. Error	&	SNR	\\
\hline															
F1	&	0.7723160979	&	0.0000031877	&	0.0026609924	&	0.0000018593	&	0.9449107704	&	0.0001113634	&	27.24290	\\
F2	&	0.4975761231	&	0.0000059818	&	0.0013999114	&	0.0000018593	&	0.3327250305	&	0.0002089767	&	113.35254	\\
F3	&	0.9988728524	&	0.0000078992	&	0.0010620928	&	0.0000018593	&	0.8358997565	&	0.0002759644	&	86.11411	\\
F4	&	1.5424135178	&	0.0000090667	&	0.0009345166	&	0.0000018593	&	0.7420582624	&	0.0003167517	&	76.43665	\\
F5	&	0.4894687433	&	0.0000083042	&	0.0010179492	&	0.0000018593	&	0.4910210062	&	0.0002901135	&	82.42263	\\
F6	&	1.0276687946	&	0.0000154359	&	0.0005344823	&	0.0000018593	&	0.4455039071	&	0.0005392634	&	43.38851	\\
F7	&	0.4407402764	&	0.0000129339	&	0.0006585206	&	0.0000018593	&	0.7890295023	&	0.0004518545	&	53.31641	\\
F8	&	0.5311150110	&	0.0000136640	&	0.0006143544	&	0.0000018593	&	0.4296130277	&	0.0004773612	&	49.75739	\\
F9	&	0.9849090641	&	0.0000124246	&	0.0006660833	&	0.0000018593	&	0.5662303913	&	0.0004340626	&	54.01709	\\
F10	&	0.7916937472	&	0.0000305106	&	0.0002807339	&	0.0000018593	&	0.3519530512	&	0.0010659082	&	22.79429	\\
F11	&	0.5795860109	&	0.0000191945	&	0.0004413958	&	0.0000018593	&	0.1273669374	&	0.0006705733	&	35.81990	\\
F12	&	1.4694682971	&	0.0000293374	&	0.0002888168	&	0.0000018593	&	0.9618532171	&	0.0010249184	&	23.61153	\\
F13	&	1.4950385433	&	0.0000348934	&	0.0002442206	&	0.0000018593	&	0.2269944994	&	0.0012190231	&	19.96318	\\
F14	&	0.3215148480	&	0.0000390646	&	0.0002173496	&	0.0000018593	&	0.2211391526	&	0.0013647462	&	17.59599	\\
F15	&	0.1311444430	&	0.0000526486	&	0.0001626092	&	0.0000018593	&	0.0198912644	&	0.0018393101	&	13.16109	\\
F16	&	2.3173994774	&	0.0000282187	&	0.0003025840	&	0.0000018593	&	0.3473188973	&	0.0009858365	&	25.19440	\\
F17	&	0.7525091100	&	0.0000240647	&	0.0003595750	&	0.0000018593	&	0.8347027269	&	0.0008407166	&	29.21993	\\
F18	&	0.5636687727	&	0.0000245231	&	0.0003419938	&	0.0000018593	&	0.8499529812	&	0.0008567298	&	27.73990	\\
F19	&	0.4080325495	&	0.0000354012	&	0.0002408583	&	0.0000018593	&	0.4065381323	&	0.0012367619	&	19.49850	\\
F20	&	0.9552365417	&	0.0000288772	&	0.0002894413	&	0.0000018593	&	0.4893944906	&	0.0010088442	&	23.47079	\\
F21	&	0.7380632474	&	0.0000274253	&	0.0003090188	&	0.0000018593	&	0.4577053105	&	0.0009581210	&	25.10511	\\
F22	&	0.1938512417	&	0.0000421370	&	0.0002021753	&	0.0000018593	&	0.8036857356	&	0.0014720830	&	16.35457	\\
F23	&	0.0994199015	&	0.0000443156	&	0.0001899003	&	0.0000018593	&	0.3254750599	&	0.0015481919	&	15.37623	\\
F24	&	0.6335700599	&	0.0000519750	&	0.0001607942	&	0.0000018593	&	0.2142595161	&	0.0018157772	&	13.05360	\\
F25	&	0.6984989836	&	0.0000382614	&	0.0002183078	&	0.0000018593	&	0.0646423964	&	0.0013366838	&	17.73914	\\
F26	&	1.1021815212	&	0.0000623869	&	0.0001233246	&	0.0000018593	&	0.4001342339	&	0.0021795241	&	10.02352	\\
F27	&	3.0871283460	&	0.0000518796	&	0.0001632645	&	0.0000018593	&	0.4609820387	&	0.0018124452	&	13.85685	\\
F28	&	0.0558343058	&	0.0000563096	&	0.0001509230	&	0.0000018593	&	0.1467764179	&	0.0019672089	&	12.21338	\\
F29	&	0.3386128997	&	0.0000446178	&	0.0001876491	&	0.0000018593	&	0.9062616530	&	0.0015587507	&	15.19642	\\
F30	&	0.9114975683	&	0.0000571347	&	0.0001503758	&	0.0000018593	&	0.6784290417	&	0.0019960341	&	12.20166	\\
F31	&	1.0588928956	&	0.0000563202	&	0.0001247546	&	0.0000018593	&	0.3458643842	&	0.0019675814	&	10.13334	\\
F32	&	0.2550253254	&	0.0000832905	&	0.0001020882	&	0.0000018593	&	0.9943019246	&	0.0029098040	&	8.25892	\\
F33	&	1.5745649357	&	0.0000806543	&	0.0001038800	&	0.0000018593	&	0.5659969012	&	0.0028177055	&	8.50920	\\
F34	&	2.3129357806	&	0.0000579092	&	0.0001470766	&	0.0000018593	&	0.7712122761	&	0.0020230923	&	12.24760	\\
F35	&	0.1567972123	&	0.0000547972	&	0.0001533881	&	0.0000018593	&	0.5794490966	&	0.0019143725	&	12.41019	\\
F36	&	1.1830227204	&	0.0000777679	&	0.0001051544	&	0.0000018593	&	0.1673114109	&	0.0027168683	&	8.55585	\\
F37	&	0.2865210885	&	0.0000739005	&	0.0001132727	&	0.0000018593	&	0.6120736605	&	0.0025817588	&	9.17030	\\
F38	&	0.8370259123	&	0.0000923417	&	0.0000929679	&	0.0000018593	&	0.5498119347	&	0.0032260127	&	7.54894	\\
F39	&	2.4799486485	&	0.0000748546	&	0.0001118372	&	0.0000018593	&	0.0476382855	&	0.0026150918	&	9.34807	\\
F40	&	0.6580805890	&	0.0000726444	&	0.0001151555	&	0.0000018593	&	0.8300884411	&	0.0025378750	&	9.35082	\\
F41	&	1.5243299827	&	0.0000565504	&	0.0001499204	&	0.0000018593	&	0.0423623742	&	0.0019756244	&	12.25617	\\
F42	&	1.9833940130	&	0.0001028949	&	0.0000814848	&	0.0000018593	&	0.0209795407	&	0.0035946974	&	6.73297	\\
F43	&	1.2758245817	&	0.0000988247	&	0.0000858415	&	0.0000018593	&	0.6400886887	&	0.0034525022	&	6.99744	\\
F44	&	1.3812585747	&	0.0001033266	&	0.0000822766	&	0.0000018593	&	0.9889746567	&	0.0036097794	&	6.71805	\\
F45	&	2.5045724493	&	0.0000942023	&	0.0000884516	&	0.0000018593	&	0.6184939074	&	0.0032910139	&	7.39540	\\
F46	&	1.9331724591	&	0.0001132792	&	0.0000732111	&	0.0000018593	&	0.1625473654	&	0.0039574785	&	6.03821	\\
F47	&	0.4563923889	&	0.0000347727	&	0.0002328754	&	0.0000018593	&	0.9428930442	&	0.0012148041	&	18.85420	\\
F48	&	1.6197884001	&	0.0001136596	&	0.0000756666	&	0.0000018593	&	0.6772422202	&	0.0039707660	&	6.20086	\\
F49	&	1.4315884040	&	0.0001219939	&	0.0000673438	&	0.0000018593	&	0.4330667564	&	0.0042619312	&	5.50389	\\
F50	&	1.0651087581	&	0.0000768301	&	0.0000654989	&	0.0000018593	&	0.8525449471	&	0.0026841078	&	5.32053	\\
\hline
\end{tabular}
\end{center}
\end{table*}

\begin{table*}
\begin{center}
\caption{The parameters obtained from the light curve analysis of KIC 2557430.}
\begin{tabular}{@{}lr@{}}
\hline
Parameter	&	Value	\\
\hline			
$q$	&	0.868$\pm$0.002	\\
$i$ ($^\circ$)	&	69.75$\pm$0.01	\\
$T_{1} (K)$	&	6913 (Fixed)	\\
$T_{2} (K)$	&	6271$\pm$1	\\
$\Omega_{1}$	&	5.8362$\pm$0.0010	\\
$\Omega_{2}$	&	5.0301$\pm$0.0009	\\
$L_{1}/L_{T}$	&	0.5576$\pm$0.0011	\\
$L_{3}/L_{T}$	&	0.0034$\pm$0.0004	\\
$g_{1}, g_{2}$	&	0.50, 0.50 (Fixed)	\\
$A_{1}, A_{2}$	&	1.0, 1.0 (Fixed)	\\
$x_{1},_{bol}, x_{2}, _{bol}$	&	0.672, 0.684 (Fixed)	\\
$x_{1}, x_{2}$	&	0.624, 0.644 (Fixed)	\\
$< r_{1} >$	&	0.2029$\pm$0.0004	\\
$< r_{2} >$	&	0.2218$\pm$0.0006	\\
\hline			
Stellar Spot Parameters for Part 1	&		\\
\hline			
$Co-Lat_{Spot I}$ $(rad)$	&	0.995$\pm$0.003	\\
$Long_{Spot I}$ $(rad)$	&	1.292$\pm$0.001	\\
$R_{Spot I}$ $(rad)$	&	0.314$\pm$0.002	\\
$T_{f Spot I}$	&	0.87$\pm$0.01	\\
$Co-Lat_{Spot II}$ $(rad)$	&	0.995$\pm$0.003	\\
$Long_{Spot II}$ $(rad)$	&	5.079$\pm$0.002	\\
$R_{Spot II}$ $(rad)$	&	0.297$\pm$0.002	\\
$T_{f Spot II}$	&	0.85$\pm$0.01	\\
\hline			
Stellar Spot Parameters for Part 2	&		\\
\hline			
$Co-Lat_{Spot I}$ $(rad)$	&	0.995$\pm$0.003	\\
$Long_{Spot I}$ $(rad)$	&	1.414$\pm$0.001	\\
$R_{Spot I}$ $(rad)$	&	0.314$\pm$0.002	\\
$T_{f Spot I}$	&	0.86$\pm$0.01	\\
$Co-Lat_{Spot II}$ $(rad)$	&	0.995$\pm$0.003	\\
$Long_{Spot II}$ $(rad)$	&	5.079$\pm$0.002	\\
$R_{Spot II}$ $(rad)$	&	0.297$\pm$0.002	\\
$T_{f Spot I}$	&	0.87$\pm$0.01	\\
\hline
\end{tabular}
\end{center}
\end{table*}

\begin{table*}
\begin{center}
\caption{All the minima times and $(O-C)_{II}$ residuals.}
\begin{tabular}{@{}rrrrrrrr@{}}
\hline
HJD (Obs)	&	$E$	&	Type	&	$(O-C)_{II}$	&	HJD (Obs)	&	$E$	&	Type	&	$(O-C)_{II}$	\\
(+24 00000)	&		&		&	(day)	&	(+24 00000)	&		&		&	(day)	\\
\hline															
55003.12760	&	37.5	&	II	&	0.00106	&	55093.96841	&	107.5	&	II	&	0.00094	\\
55003.77353	&	38.0	&	I	&	-0.00187	&	55094.61647	&	108.0	&	I	&	0.00014	\\
55004.42538	&	38.5	&	II	&	0.00111	&	55095.26539	&	108.5	&	II	&	0.00019	\\
55005.07156	&	39.0	&	I	&	-0.00157	&	55095.91399	&	109.0	&	I	&	-0.00007	\\
55005.72313	&	39.5	&	II	&	0.00114	&	55096.56293	&	109.5	&	II	&	0.00001	\\
55006.36874	&	40.0	&	I	&	-0.00211	&	55097.21191	&	110.0	&	I	&	0.00012	\\
55007.02105	&	40.5	&	II	&	0.00133	&	55097.86092	&	110.5	&	II	&	0.00027	\\
55007.66680	&	41.0	&	I	&	-0.00178	&	55098.50932	&	111.0	&	I	&	-0.00020	\\
55008.31799	&	41.5	&	II	&	0.00055	&	55099.15918	&	111.5	&	II	&	0.00081	\\
55008.96481	&	42.0	&	I	&	-0.00150	&	55099.80716	&	112.0	&	I	&	-0.00008	\\
55009.61615	&	42.5	&	II	&	0.00097	&	55101.10467	&	113.0	&	I	&	-0.00030	\\
55010.26243	&	43.0	&	I	&	-0.00161	&	55101.75495	&	113.5	&	II	&	0.00112	\\
55010.91465	&	43.5	&	II	&	0.00175	&	55102.40268	&	114.0	&	I	&	-0.00002	\\
55011.56006	&	44.0	&	I	&	-0.00170	&	55103.05152	&	114.5	&	II	&	-0.00004	\\
55012.21180	&	44.5	&	II	&	0.00117	&	55103.70005	&	115.0	&	I	&	-0.00037	\\
55012.85811	&	45.0	&	I	&	-0.00138	&	55104.34944	&	115.5	&	II	&	0.00015	\\
55013.50982	&	45.5	&	II	&	0.00146	&	55104.99768	&	116.0	&	I	&	-0.00048	\\
55014.15521	&	46.0	&	I	&	-0.00201	&	55105.64726	&	116.5	&	II	&	0.00024	\\
55016.75121	&	48.0	&	I	&	-0.00147	&	55106.29531	&	117.0	&	I	&	-0.00057	\\
55017.40293	&	48.5	&	II	&	0.00139	&	55106.94501	&	117.5	&	II	&	0.00027	\\
55018.04888	&	49.0	&	I	&	-0.00152	&	55107.59317	&	118.0	&	I	&	-0.00044	\\
55018.70095	&	49.5	&	II	&	0.00169	&	55108.24232	&	118.5	&	II	&	-0.00015	\\
55019.34706	&	50.0	&	I	&	-0.00107	&	55108.89105	&	119.0	&	I	&	-0.00029	\\
55019.99877	&	50.5	&	II	&	0.00178	&	55109.54023	&	119.5	&	II	&	0.00003	\\
55020.64480	&	51.0	&	I	&	-0.00106	&	55110.18859	&	120.0	&	I	&	-0.00047	\\
55021.29642	&	51.5	&	II	&	0.00170	&	55110.83860	&	120.5	&	II	&	0.00067	\\
55021.94223	&	52.0	&	I	&	-0.00136	&	55111.48619	&	121.0	&	I	&	-0.00060	\\
55022.59445	&	52.5	&	II	&	0.00199	&	55112.13519	&	121.5	&	II	&	-0.00047	\\
55023.23998	&	53.0	&	I	&	-0.00133	&	55112.78415	&	122.0	&	I	&	-0.00037	\\
55023.89130	&	53.5	&	II	&	0.00112	&	55113.43408	&	122.5	&	II	&	0.00070	\\
55024.53801	&	54.0	&	I	&	-0.00104	&	55114.73161	&	123.5	&	II	&	0.00050	\\
55025.18949	&	54.5	&	II	&	0.00158	&	55115.37971	&	124.0	&	I	&	-0.00026	\\
55025.83603	&	55.0	&	I	&	-0.00074	&	55116.02925	&	124.5	&	II	&	0.00041	\\
55026.48650	&	55.5	&	II	&	0.00087	&	55116.67710	&	125.0	&	I	&	-0.00060	\\
55027.13349	&	56.0	&	I	&	-0.00101	&	55117.32672	&	125.5	&	II	&	0.00015	\\
55027.78494	&	56.5	&	II	&	0.00158	&	55117.97536	&	126.0	&	I	&	-0.00007	\\
55028.43122	&	57.0	&	I	&	-0.00101	&	55118.62320	&	126.5	&	II	&	-0.00109	\\
55029.08252	&	57.5	&	II	&	0.00143	&	55119.27261	&	127.0	&	I	&	-0.00054	\\
55029.72878	&	58.0	&	I	&	-0.00117	&	55119.92215	&	127.5	&	II	&	0.00013	\\
55030.38047	&	58.5	&	II	&	0.00165	&	55120.57090	&	128.0	&	I	&	0.00002	\\
55031.02653	&	59.0	&	I	&	-0.00115	&	55121.21947	&	128.5	&	II	&	-0.00028	\\
55031.67835	&	59.5	&	II	&	0.00181	&	55121.86909	&	129.0	&	I	&	0.00048	\\
55032.32454	&	60.0	&	I	&	-0.00087	&	55122.51740	&	129.5	&	II	&	-0.00008	\\
55032.97624	&	60.5	&	II	&	0.00197	&	55123.16649	&	130.0	&	I	&	0.00015	\\
55093.31815	&	107.0	&	I	&	-0.00045	&		&		&		&		\\
\hline
\end{tabular}
\end{center}
\end{table*}

\setcounter{table}{3}
\begin{table*}
\begin{center}
\caption{All the calculated parameters of flares detected from the short cadence observational data of KIC 2557430. As it is explained in the text, the flares were separated into two groups, such as Group 1 and Group 2.}
\begin{tabular}{@{}crrrr@{}}
\hline
\hline									
Flare Time	&	$P$	&	$T_{r}$	&	$T_{d}$	&	Amplitude	\\
(+24 00000)	&	(s)	&	(s)	&	(s)	&	(Intensity)	\\
\hline									
55002.712567	&	8.039659	&	823.896576	&	5178.825504	&	0.005027	\\
55003.677739	&	0.673626	&	294.251616	&	765.056448	&	0.001400	\\
55003.837805	&	0.417511	&	176.545440	&	706.207680	&	0.000922	\\
55006.463585	&	4.793132	&	823.903488	&	3825.256320	&	0.002251	\\
55006.821862	&	3.288425	&	353.099520	&	3119.057280	&	0.003478	\\
55011.037401	&	0.402576	&	294.249888	&	706.194721	&	0.000979	\\
55014.476438	&	0.372427	&	470.802240	&	235.391616	&	0.001146	\\
55017.788094	&	0.133522	&	117.695808	&	176.552352	&	0.002310	\\
55019.178960	&	1.473707	&	176.560992	&	1294.678080	&	0.003395	\\
55020.936272	&	0.261260	&	176.542848	&	529.647552	&	0.000808	\\
55022.009048	&	0.403619	&	176.543712	&	647.350272	&	0.001168	\\
55023.572234	&	0.368507	&	353.094336	&	529.637184	&	0.000931	\\
55023.594030	&	0.230196	&	294.247296	&	294.246433	&	0.001103	\\
55024.469959	&	0.277295	&	235.399392	&	353.093472	&	0.001261	\\
55024.474727	&	0.140336	&	58.847904	&	117.703584	&	0.001645	\\
55024.887489	&	20.159287	&	765.035712	&	7532.702784	&	0.009431	\\
55025.677595	&	0.223677	&	117.703584	&	411.941376	&	0.000667	\\
55026.362807	&	0.265899	&	176.560128	&	470.788416	&	0.000929	\\
55027.526849	&	6.002558	&	117.703584	&	3589.797312	&	0.007432	\\
55028.252247	&	0.496766	&	647.338176	&	294.245568	&	0.001938	\\
55029.538208	&	0.297052	&	588.481632	&	117.694944	&	0.000833	\\
55029.585206	&	0.257707	&	117.703584	&	647.337312	&	0.000809	\\
55030.074933	&	0.490036	&	117.703584	&	765.032256	&	0.001896	\\
55031.182439	&	37.020060	&	647.337312	&	9180.440640	&	0.015910	\\
55031.294143	&	1.377747	&	470.795328	&	1294.665984	&	0.002374	\\
55094.907384	&	0.306107	&	470.773728	&	294.223104	&	0.001201	\\
55094.922368	&	0.249014	&	470.765952	&	294.239520	&	0.000799	\\
55094.930541	&	0.247567	&	411.929280	&	294.223104	&	0.000698	\\
55095.452257	&	6.524747	&	470.765088	&	3883.862304	&	0.005723	\\
55097.401539	&	0.253873	&	470.773728	&	294.230880	&	0.001011	\\
55099.405307	&	0.332166	&	647.307072	&	176.533345	&	0.001022	\\
55100.654425	&	1.146446	&	176.541984	&	1412.309952	&	0.002284	\\
55101.914441	&	0.663007	&	470.763360	&	529.617312	&	0.001315	\\
55102.730386	&	6.538134	&	411.918912	&	5296.149793	&	0.003590	\\
55105.101255	&	0.438149	&	882.690048	&	117.688896	&	0.001212	\\
55107.150650	&	3.017127	&	941.534496	&	2353.840128	&	0.003184	\\
55107.706417	&	1.475751	&	353.074464	&	1706.544288	&	0.002258	\\
55113.089730	&	0.258550	&	117.688032	&	294.229152	&	0.001755	\\
55122.376337	&	11.891936	&	706.154111	&	6414.192577	&	0.005768	\\
55123.516474	&	0.211945	&	176.540256	&	235.385568	&	0.001146	\\
\hline
\end{tabular}
\end{center}
\end{table*}

\setcounter{table}{3}
\begin{table*}
\begin{center}
\caption{Continued From Previous Page.}
\begin{tabular}{@{}crrrr@{}}
\hline
Flare Time	&	$P$	&	$T_{r}$	&	$T_{d}$	&	Amplitude	\\
(+24 00000)	&	(s)	&	(s)	&	(s)	&	(Intensity)	\\
\hline									
55002.927807	&	1.532509	&	1588.961664	&	1647.801792	&	0.001251	\\
55003.850747	&	0.248598	&	411.956064	&	706.207680	&	0.000994	\\
55004.935797	&	2.045424	&	1471.254624	&	3413.312352	&	0.001144	\\
55005.242309	&	0.579122	&	647.340768	&	941.601024	&	0.000969	\\
55006.693808	&	0.862105	&	1412.395488	&	765.063360	&	0.001034	\\
55006.874991	&	0.939750	&	1471.252896	&	353.098656	&	0.001262	\\
55007.902142	&	0.913513	&	411.946560	&	1647.795744	&	0.001302	\\
55008.788298	&	1.837626	&	765.053855	&	3119.044320	&	0.001407	\\
55008.830528	&	0.338779	&	529.651872	&	529.651872	&	0.000985	\\
55010.425743	&	3.496779	&	765.043488	&	5061.092544	&	0.001656	\\
55010.625997	&	1.548999	&	1824.355008	&	1765.497600	&	0.001478	\\
55011.703549	&	0.394312	&	765.052128	&	706.194720	&	0.000081	\\
55013.111450	&	7.063702	&	2177.448480	&	7650.476352	&	0.002805	\\
55013.274922	&	0.771074	&	882.755712	&	1353.540672	&	0.001247	\\
55020.351864	&	1.142096	&	529.646688	&	2118.580704	&	0.001028	\\
55025.701434	&	0.374385	&	1118.137824	&	58.847904	&	0.000663	\\
55025.987507	&	0.485702	&	1118.145600	&	765.034848	&	0.000866	\\
55029.357710	&	0.482718	&	588.491136	&	1000.439424	&	0.001043	\\
55033.127042	&	5.713145	&	2059.712064	&	7356.103776	&	0.001792	\\
55093.599688	&	0.145763	&	353.077056	&	235.395072	&	0.000677	\\
55093.609905	&	0.191866	&	58.845312	&	765.005472	&	0.000480	\\
55096.304983	&	5.814991	&	1294.614144	&	8356.175136	&	0.001964	\\
55097.761155	&	0.766162	&	470.773728	&	1294.613280	&	0.001038	\\
55099.423015	&	0.122277	&	58.844448	&	529.608672	&	0.000760	\\
55101.335515	&	2.580089	&	3059.986464	&	3530.776608	&	0.001173	\\
55105.104661	&	0.147698	&	294.230880	&	411.926688	&	0.000608	\\
55108.021080	&	0.722516	&	176.532480	&	1412.305632	&	0.001171	\\
55118.921871	&	0.751852	&	1235.760192	&	1412.292672	&	0.000927	\\
55119.738493		&	0.989688	&	941.531040	&	1412.301312	&	0.000948	\\
\hline
\end{tabular}
\end{center}
\end{table*}

\setcounter{table}{4}
\begin{table*}
\begin{center}
\caption{The parameters obtained from the OPEA models, using the least-squares method.}
\begin{tabular}{@{}lr@{}}
\hline
Parameters of the OPEA for	&		\\
\hline			
$y_{0}$	&	-0.9722$\pm$0.0518	\\
$Plateau$	&	1.4336$\pm$0.1104	\\
$K$	&	0.00030426$\pm$0.00003665	\\
$Tau$	&	3286.6	\\
$Half-life$	&	2278.1	\\
$Span$	&	2.4058$\pm$0.0888	\\
\hline			
95$\%$Confidence Intervals	&		\\
\hline			
$y_{0}$	&	-1.0771 to -0.8673	\\
$Plateau$	&	1.2099 to 1.6573	\\
$K$	&	0.00022996 to 0.00037857	\\
$Tau$	&	2641.5 to 4348.5	\\
$Half-life$	&	1831.0 to 3014.2	\\
$Span$	&	2.2258 to 2.5858	\\
\hline			
Goodness of Fit	&		\\
\hline			
$R^{2}$	&	0.9721	\\
\hline			
p-values	&		\\
\hline			
D'Agostino \& Pearson	&	0.002	\\
Shapiro-Wilk	&	0.004	\\
Kolmogorov-Smirnov	&	0.001	\\
	&		\\
\hline
Parameters of the OPEA for Group 2	&		\\
\hline			
$y_{0}$	&	-1.2056$\pm$0.0902	\\
$Plateau$	&	0.7550$\pm$0.0677	\\
$K$	&	0.00038271$\pm$0.000043703	\\
$Tau$	&	2613	\\
$Half-life$	&	1811.2	\\
$Span$	&	1.9606$\pm$0.0835	\\
\hline			
95$\%$Confidence Intervals	&		\\
\hline			
$y_{0}$	&	-1.3910 to -1.0202	\\
$Plateau$	&	0.6159 to 0.8941	\\
$K$	&	0.00029285 to 0.00047256	\\
$Tau$	&	2116.1 to 3414.7	\\
$Half-life$	&	1466.8 to 2366.9	\\
$Span$	&	1.7891 to 2.1322	\\
\hline			
Goodness of Fit	&		\\
\hline			
$R^{2}$	&	0.9575	\\
\hline			
p-values	&		\\
\hline			
D'Agostino \& Pearson	&	0.008	\\
Shapiro-Wilk	&	0.009	\\
Kolmogorov-Smirnov	&	0.001	\\
\hline
\end{tabular}
\end{center}
\end{table*}

\setcounter{table}{5}
\begin{table*}
\begin{center}
\caption{Flare frequencies computed for all flares and grouped flares.}
\begin{tabular}{@{}lccc@{}}
\hline
Parameters	&	All	&	Group 1	&	Group 2	\\
\hline							
Total Time (h)	&	1467.22299	&	1467.22299	&	1467.22299	\\
Flare Number	&	69	&	40	&	29	\\
Total Equivalient Duration (s)	&	164.42435	&	121.42111	&	43.00324	\\
$N_{1}$ ($h^{-1}$)	&	0.04703	&	0.02726	&	0.01977	\\
$N_{2}$	&	0.00003	&	0.00002	&	0.00001	\\
\hline
\end{tabular}
\end{center}
\end{table*}

\end{document}